\documentclass[aps,preprint,nofootinbib,preprintnumbers,eqsecnum,superscriptaddress]{revtex4}
\pdfoutput=1

\usepackage{color}

\usepackage[
      colorlinks=true,
      linkcolor=blue,
      urlcolor=blue,
      filecolor=black,
      citecolor=red,
      pdfstartview=FitV,
      pdftitle={},
        pdfauthor={Martin Ammon, Alejandra Castro, Nabil Iqbal},
        pdfsubject={},
        pdfkeywords={},
        pdfpagemode=None,
        bookmarksopen=true
      ]{hyperref}

\usepackage[font=footnotesize,labelfont=bf,justification=centerlast,width=.94\textwidth]{caption}

\usepackage[normalem]{ulem}
\usepackage{amsmath}
\usepackage{enumerate}
\usepackage{amsfonts}
\usepackage{yfonts}

\usepackage{subfigure}
\usepackage{psfrag}

\usepackage{epsfig}
\usepackage[latin1]{inputenc}
\usepackage{float}
\usepackage{graphicx}
\usepackage{cancel}
\usepackage{mathrsfs}
\usepackage{amssymb}
\usepackage{amsfonts}
\usepackage{amsmath}
\usepackage{slashed}

\newcommand \tr {\mbox{{\bf Tr}}}

\usepackage{graphicx}
\usepackage{bm}

\def\({\left(}
\def\){\right)}
\def\[{\left[}
\def\]{\right]}
\def\<{\langle}
\def\>{\rangle}

\newcommand{\bmat}{\begin{bmatrix}}
\newcommand{\emat}{\end{bmatrix}}

\def\Tr{\mathop{\rm Tr}}
\def\tr{\mathop{\rm tr}}

\newcommand\half{{\ensuremath{\frac{1}{2}}}}
\newcommand\p{\ensuremath{\partial}}

\newcommand\field[1]{{\ensuremath{\mathbb{{#1}}}}}

\newcommand{\RR}{\field{R}}

\newcommand{\be}{\begin{equation}}
\newcommand{\ee}{\end{equation}}
\newcommand{\bea}{\begin{eqnarray}}
\newcommand{\eea}{\end{eqnarray}}
\newcommand{\bwt}{\begin{widetext}}
\newcommand{\ewt}{\end{widetext}}

\newcommand{\bi}{\begin{itemize}}
\newcommand{\ei}{\end{itemize}}
\newcommand{\ben}{\begin{enumerate}}
\newcommand{\een}{\end{enumerate}}
\newcommand{\bca}{\begin{cases}}
\newcommand{\eca}{\end{cases}}
\newcommand{\bln}{\begin{align}}
\newcommand{\eln}{\end{align}}
\newcommand{\bst}{\begin{split}}
\newcommand{\est}{\end{split}}

\newcommand\al{{\alpha}}
\newcommand\ep{\epsilon}
\newcommand\sig{\sigma}

\newcommand\lam{\lambda}

\newcommand\om{\omega}

\def\th{{\theta}}

\newcommand\ha{{\half}}

\def\le{\left}
\def\ri{\right}

\newcommand\sA{{\ensuremath{{\mathcal A}}}}

\newcommand\sD{{\ensuremath{{\mathcal D}}}}

\newcommand\sL{{\ensuremath{{\mathcal L}}}}

\newcommand\sO{{\ensuremath{{\mathcal O}}}}
\newcommand\sR{{\ensuremath{{\mathcal R}}}}

\newcommand\sW{{\mathcal W}}
\newcommand\sJ{{\mathcal J}}

\newcommand\bpsi{{\bar \psi}}

\newcommand{\slt}{SL(2,\RR)}
\newcommand{\slh}{SL(3,\RR)}

\newcommand{\bA}{\bar{A}}

\newcommand{\ba}{\bar{a}}
\newcommand{\bchi}{\bar{\chi}}

\begin{document}

\title {Wilson Lines and Entanglement Entropy in Higher Spin Gravity}

\preprint{NSF-KITP-13-112}

\author{Martin Ammon}
\affiliation{Theoretisch-Physikalisches Institut, Friedrich-Schiller Universit\"at Jena, Max-Wien-Platz 1, D-07743 Jena, Germany}
\affiliation{Department of Physics and Astronomy, University of California, Los Angeles, CA 90095, USA}

\author{Alejandra Castro}
\affiliation{Center for the Fundamental Laws of Nature, Harvard University, Cambridge, MA 02138, USA}

\author{Nabil Iqbal}
\affiliation{Kavli Institute for Theoretical Physics, University of California, Santa Barbara CA 93106  }

\begin{abstract}

Holographic entanglement entropy provides a direct connection between classical geometry and quantum entanglement; however the usual prescription does not apply to theories of higher spin gravity, where standard notions of geometry are no longer gauge invariant. We present a proposal for the holographic computation of entanglement entropy in field theories dual to higher spin theories of gravity in AdS$_3$. These theories have a Chern-Simons description, and our proposal involves a Wilson line in an infinite-dimensional representation of the bulk gauge group. In the case of spin$-2$ gravity such Wilson lines are the natural coupling of a heavy point particle to gravity and so are equivalent to the usual prescription of Ryu and Takayanagi. For higher spin gravity they provide a natural generalization of these ideas. We work out spin$-3$ gravity in detail, showing that our proposal recovers many expected results and computes thermal entropies of black holes with higher spin charge, finding agreement with previous expressions in 
the literature. We encounter some peculiarities in the case of non-unitary RG flow backgrounds and outline future generalizations. 
\end{abstract}

\vfill

\today

\maketitle

\tableofcontents

\section{Introduction}
The emergence of geometry from a quantum field theory is becoming a tractable quest in the era of AdS/CFT. A variety of quantum field theories are dual in an appropriate regime to theories of classical gravity.  The fundamental mechanism of the duality is  not fully understood, nevertheless we expect that the geometric data characterizing the gravitational side of the duality can be viewed as ``emergent'',  hence it can be reconstructed from appropriate observables in the dual field-theoretical formulation. 

Remarkably, there is a universal field-theoretical observable that provides a direct probe of the geometry of the bulk dual. The entanglement entropy of a subsystem $X$ is an observable that measures the degree of quantum entanglement between $X$ and the rest of the system, and can be defined for any quantum mechanical system \cite{Holzhey:1994we, Calabrese:2004eu,CardyCFT}. A conjecture by Ryu and Takayanagi states that if a field theory has a ``simple'' Einstein gravity dual,\footnote{To be precise, ``simple'' here denotes the Einstein-Hilbert action coupled minimally to matter, and mild deviations thereof.} then the entanglement entropy of a geometric subregion $X$ in the field theory is given by the area of a minimal surface that ends on the boundary at $\p X$ \cite{Ryu:2006bv,Ryu:2006ef}. This is a beautiful statement that relates two very fundamental ideas on the two sides of the duality: classical geometry in the bulk and quantum entanglement on the boundary. See \cite{Nishioka:2009un,Takayanagi:2012kg} for recent reviews and a complete set of references on this topic. 

 There exist as well examples of AdS/CFT  where the usual notion of classical geometry on the gravitational side of the duality is not clear -- in the language used above, they are not ``simple.'' Theories of higher spin gravity fall into this class. In addition to the bulk metric, these theories possess a (possibly infinite number of) fields with spin greater than two that interact nonlinearly with each other and with the graviton \cite{Vasiliev:2001ur,Bekaert:2005vh}. Fascinatingly, these theories also possess enlarged gauge redundancies that act nontrivially and unfamiliarly on the metric. Black holes in such theories can be gauge transformed into traversable wormholes; causal structures can be changed and singularities removed. Thus the very notion of ``geometry'' in these theories is not gauge invariant and must be replaced by some other more general concept. It is simultaneously exciting and confusing to imagine what such a concept might be. 

On the other hand, these higher spin theories of gravity are thought to be dual to perfectly ordinary CFTs; thus the notion of the {\it entanglement entropy} of the dual field theory is entirely well defined. There is now a very natural question: what is the bulk object that computes entanglement entropy in the boundary theory? The conventional Ryu-Takayanagi prescription clearly requires modification, as the idea of a proper distance is no longer meaningful. If such an object could be found in higher spin theory, it would be tempting to think of it as defining a new, generalized notion of geometry, one that takes as fundamental the notion of the entanglement of the dual field theory.

In this work we initiate an investigation into these issues. We will work within the simplest possible theory of higher spin gravity, that containing a single extra spin $3$ field in a 3d bulk. We make a concrete proposal for a bulk object that computes entanglement entropy and apply this prescription to various solutions of higher spin gravity. 

\subsection{Entanglement entropy from Wilson lines}
A key fact that makes our analysis possible is that higher-spin theories in 3d admit a simple Lagrangian description in the bulk: they can be written in terms of a doubled Chern-Simons theory with gauge group $SL(N, \mathbb{R})$. The case $N = 2$ is ordinary Einstein gravity in AdS$_3$; for $N > 2$ we find a higher spin theory. We seek a formula for entanglement entropy in terms of the data specifying a classical solution to the Chern-Simons theory, i.e. gauge connections $A$, $\bA$ valued in $SL(N, \mathbb{R})$. We point out that even in the case $N = 2$, where we have ordinary Einstein gravity and we know the answer should simply be the length of a spatial geodesic, this is not trivial: the Chern-Simons representation obscures many aspects of a geometric interpretation.

\begin{figure}[h]
\includegraphics[width=0.6\textwidth]{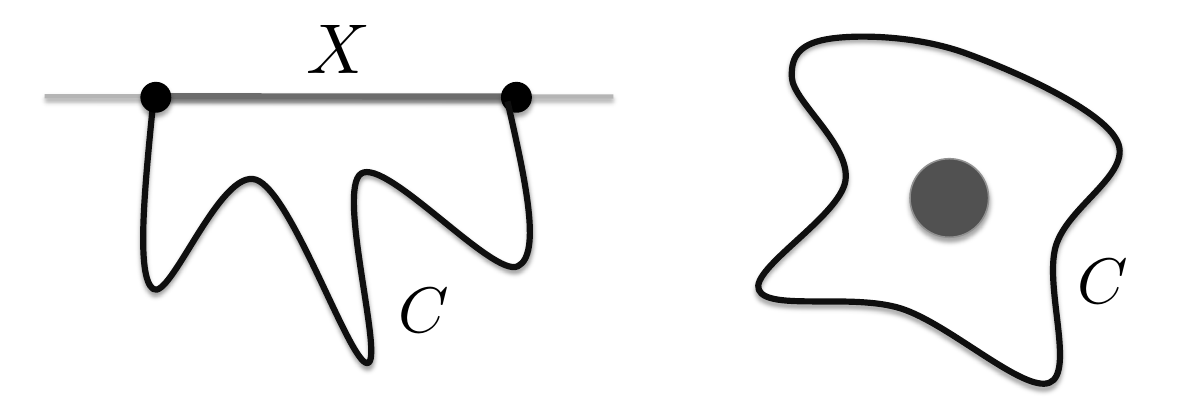}
\caption{{\it Left:} Entanglement entropy of an interval $X$ is found by computing a particular Wilson line along a curve $C$; the line ends on the boundary at $\p X$. {\it Right:} The thermal entropy of a black hole can be found by evaluating a closed Wilson loop around the horizon. }
\label{fig:openandclosed}
\end{figure}

We propose that the entanglement entropy for a single interval $X$ is
\be
S_{\rm EE} = -\log\le(W_{\sR}(C)\ri)~, \label{EEform}
\ee 
where $W_{\sR}(C)$ is a bulk Wilson line in a representation $\sR$ of the gauge group along a curve $C$ that ends at the boundary at $\p X$, i.e. as in Figure \ref{fig:openandclosed}. The choice of representation $\sR$ is crucial: it is an {\it infinite-dimensional} highest-weight representation of $SL(N, \mathbb{R})$. To define the representation we need to specify its Casimirs: we relate the quadratic Casimir to the central charge of the theory and argue that all higher-order Casimirs are zero.\footnote{The Casimirs do not specify uniquely the representation ${\sR}$; additional data is needed.  However for the cases studied here it will suffice to specify only this data of the representation.} 

This formula may appear somewhat strange:  in \eqref{EEform}, the actual bulk path $C$ taken does not matter (even for $N=2$). The only relevant data of the curve is the location of the endpoints, and whether or not the background contains a black hole (see Figure \ref{fig:openandclosed}). However in the case of $N =2$ gravity this Wilson line is actually the natural coupling of a massive point particle to AdS$_3$ gravity, a connection that we review in section \ref{sec:probeSL2}. It should then not be surprising that it computes the length of a bulk geodesic and is thus equivalent to the Ryu-Takayanagi prescription. For $N > 2$ it generalizes these ideas in a manner that is manifestly invariant under higher-spin gauge symmetry. 

We do not feel that we have really {\it proven} that this proposal is correct. Rather it should be thought of as a conjecture motivated by the following:
\ben
\item This is the natural generalization of the idea of a proper distance in a sense that we expound upon at length.
\item Under some reasonable assumptions---the same that appear when trying to justify the Ryu-Takayanagi prescription---this object implements the replica trick approach to computing entanglement entropy. 
\een

It may seem daunting to compute a trace in an infinite-dimensional representation. Following \cite{Witten:1989sx,Carlip:1989nz}, we construct this representation as the Hilbert space of an auxiliary quantum-mechanical system described by the path integral of a field $U$ that lives on the Wilson line and couples to the bulk gauge connections $A$, $\bA$. In an appropriate classical limit the on-shell action of $U$ computes the Wilson line and the problem reduces to solving its classical equations of motion. It turns out as the bulk gauge connections are flat, this can be done explicitly and the on-shell action can be algebraically related to data defining the bulk geometry.  

An important consistency check on this formula is that when the curve $C$ is a closed loop enclosing, for example, a black hole horizon equation \eqref{EEform} should compute the thermal entropy of the black hole. We check this, finding consistency with previous results in the literature (although these results themselves are not free from controversy, as we will discuss). 

\subsection{Summary of paper}
Here we present a brief summary of the paper. The first half of this paper deals with ordinary spin-$2$ gravity on AdS$_3$, written in the $\slt$ Chern-Simons representation. In Section \ref{sec:probeSL2} we introduce the Wilson line that is our focus and develop the technology to evaluate an infinite-dimensional trace using an auxiliary field $U(s)$. We discuss the connection between this Wilson line and an ordinary massive particle probe. We show that the bulk geodesic equation (with respect to a metric formulation of the theory) makes a somewhat unexpected appearance and can be used to show that this Wilson line computes proper distances. In Section \ref{sec:entgrav} we review the connection between this probe and entanglement entropy, arguing that it implements the replica trick approach to computing entanglement entropy. We also develop some tools for computing this Wilson line using algebraic techniques involving $\slt$ gauge invariance, reproducing known results from 2d CFT without requiring the 
solution of any differential equations. We note that these first three sections make no reference to higher spin gravity but may still be of interest for their own sake, demonstrating a new way to think about the Ryu-Takayanagi formula. 

In Section \ref{sec:probeSL3} we finally turn to higher spin gravity, written as a $\slh$ Chern-Simons theory. We show that the Wilson line construction generalizes naturally to higher spin gravity, with some new free parameters which we fix by appealing to the replica trick. In Section \ref{sec:thermhs} we evaluate closed Wilson loops around black hole horizons and demonstrate that they compute black hole entropy. In Section \ref{sec:EEhs} we evaluate the entanglement entropy of an open interval in various higher spin backgrounds. While for the most part our results are sensible, we encounter some peculiarities in the case of a non-unitary RG flow. We conclude with a summary of important results and a list of future directions in Section \ref{sec:conc}. 

While this work was in progress we came to know of \cite{JJ}, which also studies the relation between Wilson lines and entanglement entropy in higher spin theories of gravity. Though our initial starting point is quite different, our results for open Wilson lines, in the limit where the end points are at the boundary, are in agreement with the expressions presented in \cite{JJ}. 

\section{Massive probes and Wilson lines in AdS$_3$ gravity} \label{sec:probeSL2}

Our goal in this section is to write down a version of the Ryu-Takayanagi formula using language that is natural to the Chern-Simons formulation of 3d gravity. It seems clear that the relevant construction should have something to do with bulk Wilson lines as we  will justify appropriately. 

The derivations carried out in this section are not novel, even in the context of 3d gravity \cite{Witten:1988hf}. For instance,  \cite{Witten:1989sx,Carlip:1989nz,Skagerstam:1989ti,deSousaGerbert:1990yp,Vaz:1994tm} discuss  in some depth how to interpret  the Wilson loop operator as the effective action of a massive probe. Our aim here is to review these derivations and recast them without reference to local (metric-like) fields. This will allow us to give a different derivation of a geodesic distance which makes it possible to compute entanglement entropy in theories where the  Ryu-Takayanagi  proposal is not applicable. 

\subsection{AdS$_3$ gravity as a Chern-Simons theory} \label{sec:sl2grav}

We start by reviewing some aspects of 3d gravity.  It is well known that 3d general relativity has no propagating degrees of freedom: from the bulk point of view it is purely topological and so can be cast as a Chern-Simons theory \cite{Achucarro:1987vz,Witten:1988hc}. We will be interested in the case of AdS$_3$ gravity, where the relevant Chern-Simons gauge group is $G=SO(2,2)$. The Einstein-Hilbert action can be written as
\bea
S_{\rm EH} [e,\omega]&=& S_{CS}[{\cal A}] \cr &=& \frac{k}{4\pi}\int_{\cal M} \Tr\left({\cal A} \wedge d{\cal A} + \frac{2}{3} {\cal A} \wedge {\cal A} \wedge {\cal A}\right)~,
\eea
with ${\cal A}\in  so(2,2)$.  
Here $k$ is the level of the Chern-Simons theory, and $\cal M$ is a 3-manifold with topology $\RR\times D^2$ or $\RR^{1,2}$. The relation to the conventional gravitational vielbein and spin connection is
\be
{\cal A}_i= e^a_i P_a + \omega^a_i M_a~,
\ee
 where $M_a$ are Lorentz generators and $P_a$ are  translations in $so(2,2)$. Under an infinitesimal gauge transformation $\Lambda=\rho^a P_a +\tau^a M_a$ we have
 \be
 \delta {\cal A}_i= \partial_i \Lambda +[{\cal A}_i,\Lambda]~,
 \ee
 which gives the transformation laws
 \bea
 \delta e^a_i&=&\partial_i \rho^a+\epsilon^{abc}e_{i\, b}\tau_c +\epsilon^{abc}\omega_{i\, b}\rho_c~,\cr
 \delta \omega^a_i&=&\partial_i \tau^a+\epsilon^{abc}\omega_{i\, b}\tau_c+{1\over \ell^2}\epsilon^{abc}e_{i\, b}  \rho_c~.
 \eea

 It will be convenient to write the gauge group $SO(2,2)$ as $SL(2,\RR)\times SL(2,\RR)$. The flat connection ${\cal A}$ can then be decomposed as two pairs of connections 
\be
A= (\omega^a +{1\over \ell} e^a)J_a^+ ~,\quad  \bar A= (\omega^a -{1\over \ell}e^a)J_a^-~, \label{vbspin}
\ee
with $J_a^\pm={1\over 2}(M_a\pm \ell P_a)$. Here $\ell$ is the AdS radius, and Newton's constant is related to the Chern-Simons level via
  \be\label{eq:level}
  k={\ell\over 4G_3}~.
  \ee
We will  denote the generators of $sl(2,\mathbb{R})$ simply as $J_a$. After performing this decomposition the action can be written 
\be
S_{\rm EH} = S_{CS}[A] - S_{CS}[\bA]~,
\ee
where the trace operation used in defining the Chern-Simons form is now the usual bilinear form on the $sl(2,\mathbb{R})$ Lie algebra. We will usually set the AdS radius $\ell = 1$. The metric can be recovered as
\be
g_{\mu\nu} =2{\rm tr}_f(e_{\mu} e_{\nu}) ~,\label{metsl2}
\ee
where we are taking the trace in the fundamental representation. Further details of our conventions can be found in appendix \ref{app:conv}. 

\subsection{Wilson line and a massive point particle}
 In this subsection we will revisit the physical interpretation of Wilson lines in AdS$_3$ gravity. To that end, we start by introducing the Wilson line operator:
\be\label{eq:WRC}
W_{\cal R}(C)={\rm tr}_{\cal R} \left( {\cal P}\exp \int_C {\cal A}\right)~.
\ee
Here $\cal R$ is a representation of the gauge group $G$, and $C$ is a curve on $\cal M$.  In particular, if the path $C$ is closed,  the Wilson loop is invariant under 
\be
{\cal A} \to {\cal A}' = \Lambda^{-1}{\cal  A} \Lambda + \Lambda^{-1} d\Lambda~,
\ee
with $\Lambda$ a globally defined gauge parameter. In our particular application we have $G = \slt \times \slt$, but much of our discussion will be more general. 

Our goal is to construct via a Wilson line a probe of the geometry that achieves the same goal as a geodesic, in that the probe will report back a number which is the {\it proper distance} in the bulk, but will do so using the gauge connections $A$ and $\bar{A}$ rather than the bulk metric itself. A geodesic can be understood as the trajectory followed by a dynamical massive point particle in the 3d bulk. Thus our question is essentially equivalent to asking: how does one couple a massive particle to 3d gravity in the Chern-Simons formulation? This problem has been studied extensively in the context of asymptotically flat 3d gravity \cite{Witten:1989sx,Carlip:1989nz}, and in this section we will essentially rephrase much of that discussion in the context of AdS$_3$ gravity. 

Besides the choice of curve $C$, the Wilson line depends on the representation $\sR$. We seek a representation that carries the data of a massive particle in AdS$_3$.  One thought might be to use a familiar finite-dimensional representation, say the fundamental or adjoint of $\slt$. This cannot be entirely correct. A massive particle moving in the bulk is characterized by (at least) two numbers, its mass $m$ and spin $s$. While in a quantum theory there may be some quantization condition on the mass or spin, in some classical limit they should be continuously tunable. This choice of parameter will clearly affect the backreaction on the geometry and thus should be encoded in our choice of representation; however, in ordinary finite-dimensional representations there is no natural place for these numbers to sit. Furthermore, and perhaps more fundamentally, there are no unitary finite-dimensional representations of $\slt$, suggesting that in a fully quantum theory this can not be the correct choice. 

The natural unitary representations for the massive probe are {\it infinite-dimensional}. Consider then the highest-weight representation of $\slt$, defined in the standard way via a highest-weight state $|h\rangle$ which satisfies
\be\label{eq:lwrep}
J_1 |h \rangle = 0~, \qquad J_0 |h\rangle = h | h \rangle~.
\ee 
There is an infinite tower of descendants found by acting with the raising operator $|h, n \rangle \sim (J_{-1})^n |h \rangle$: these form an irreducible, unitary, and infinite-dimensional representation of $\slt$.\footnote{There are of course additional infinite dimensional representation of $\slt$, and it might be interesting to understand their physical interpretation in the present context. For the purposes of our discussion it is sufficient to just consider the highest-weight representation. } They can be conveniently labeled by the value of the quadratic Casimir of the algebra,
\bea\label{eq:c2rep}
C_2& =& 2J_0(J_0 - 1) - 2 J_{-1}J_1\cr
&=& 2 J_0^2 - (J_{-1}J_1 + J_{1}J_{-1})~,
\eea
which commutes with all the elements of the algebra and thus is a constant on the representation. We may evaluate it on the highest-weight state to find $C_2 = 2h(h-1)$. This number, or equivalently the energy $h$ of the highest-weight state, is a parameter that labels the representation. 

We claim that the choice of representation for a Wilson line that naturally corresponds to a massive particle moving in the AdS$_3$ bulk is the {\it infinite-dimensional highest-weight representation of $\slt \times \slt$, characterized by $(h, \bar{h})$}. In the context of AdS$_3$/CFT$_2$ this seems particularly transparent: clearly $(h, \bar{h})$ correspond to the conformal dimensions of the dual CFT operator. In this notation $h+\bar h$ is related to the bulk mass and $h - \bar{h}$ determines the bulk spin;  the usual dictionary in AdS/CFT.

We now turn to the issue of how to work with such an infinite-dimensional representation. This can be done following \cite{Witten:1989sx}. Note first that infinite-dimensional representations of symmetry algebras are common in physics: they are the Hilbert spaces of quantum mechanical systems. We will generate the states by constructing an auxiliary quantum mechanical system that lives on the Wilson line. This auxiliary system can be constructed as a path integral over some fields $U$ which have a global symmetry group $G$: we will pick the dynamics of $U$ so that upon quantization the Hilbert space of the system will be precisely the representation $\sR$ that we want. Now the global symmetries of this system will be coupled to the external gauge fields $\sA$ in the obvious way. The trace over $\sR$ can then be taken to be the usual trace over the Hilbert space; in particular, for a closed Wilson loop $W_{\cal R}(C)$ is nothing more than the partition function of the quantum mechanical system, where $\sA$ 
can now be identified as a contribution to an evolution operator around $C$. 

To be slightly more concrete, we can replace the trace over the Hilbert space by the appropriate path integral.  So we have
\be\label{aa:pathint}
W_{\cal R}(C)= \int {\cal D}U \exp[-S(U; {\cal A})_C] ~,
\ee
and in general we will decompose the action $S(U;A)_C$ as 
\be
S(U;{\cal A})_C = S(U)_{C,\rm free} + S(U;{\cal A})_{C, \rm int}~.
\ee
The action $S(U)_{C,\rm free}$ has $G$ as a global symmetry;  $S(U,{\cal A})_{C,\rm int}$ will promote this global symmetry to a gauge symmetry along the worldline by using the pullback of the bulk gauge fields to the worldline. 

We have expressed \eqref{aa:pathint} as a Euclidean path integral, where $S(U; {\cal A})_C$ is real, and hence the contribution of $W_{\cal R}(C)$ to the full Chern-Simons path integral is exponentially suppressed. This will be convenient for later purposes, since we will see that the contributions to the path integral that are relevant for evaluating gravitational entropy  (a la \cite{Lewkowycz:2013nqa}) will come from solutions to the Euclidean equations of motion; in some sense they are the analog of instanton effects in quantum mechanics. In the Lorentzian theory there would be a factor of ``$i$'' in \eqref{aa:pathint}.

We emphasize that our construction of an effective action will not be unique, in the sense that there are many choices of probes and effective actions that we could attach to them. But this should not affect the evaluation of \eqref{aa:pathint}; the Wilson loop should only be sensitive to choices of  $\cal R$ and $C$.

\subsection{Constructing a topological probe}\label{sec:ctp}

In this section we will implement the logic of the previous section and construct a version of the probe $U$.\footnote{See appendix \ref{app:carlip} for a summary  of the construction done in \cite{Witten:1988hf,Witten:1989sx,Carlip:1989nz} which uses a more geometrical implementation.} We will devise an explicit system with the appropriate global symmetry and then we will couple it to the external gauge fields. 

Take $U$ to live in the group manifold $\slt$. An appropriate system is described by the following first-order action \cite{Dzhordzhadze:1994np}:
\be
S(U,P)_{\rm free} = \int_C ds \le( \Tr\le(P U^{-1} \frac{dU}{ds}\ri) + \lam(s)\le(\Tr(P^2) - c_2\ri)\ri)~. \label{ac1}
\ee
$P$ is a canonical momentum conjugate to $U$, and lives in the Lie algebra $sl(2,\mathbb{R})$.  The variable $s$ parametrizes the curve $C$; for concreteness, we take $s \in [0, s_f]$. $\lam(s)$ is a Lagrange multiplier that constrains the norm of $P$, and $c_2$ will turn out to be the value of the Casimir characterizing the representation. Here we use `$\Tr$' as a short cut for a contraction using the Lie algebra metric $\delta_{ab}$, i.e. for $P=P_aJ^a$ we have\footnote{We emphasize that `$\Tr$' is not the matrix trace; we will use the symbol `$\tr_{\cal R}$' to denote traces over a matrix representation ${\cal R}$.}
\be
\Tr (P^2) = P_a P_b \delta^{a b}=2P_0^2-(P_{-1}P_1+P_1P_{-1})~. 
\ee

This action has a global symmetry group $\slt \times \slt$, where the two copies of $\slt$ act from the left and the right on $U$:
\be\label{eq:UG}
U(s)\to L U(s)R ~,\quad  P \to R^{-1} P(s) R ~, \qquad {L,R}\in SL(2,\RR)~.
\ee
Since $P$ transforms only under $R$ it is actually a ``right'' momentum $P_R$; we will often omit the subscript $R$ for brevity, i.e. $P_R \equiv P$. The system could also have been formulated in terms of a ``left'' momentum $P_L$ if the initial kinetic term had been instead written $\Tr\le(\frac{dU}{ds}U^{-1} P_L\ri)$. In the current formulation $P_L$ is related to $P_R$ as $P_L = U P_R U^{-1}$. $P_L$ and $P_R$ are both conserved quantities which arise as Noether currents  associated with the left and right global symmetries. 

The equations of motion are
\be
U^{-1} \frac{dU}{ds} + 2 \lam P = 0~, \qquad \frac{dP}{ds} = 0~, \qquad \Tr P^2 = c_2 \ . \label{eom1}
\ee
The canonical structure of this system is slightly non-standard due to the nontrivial kinetic term. The Poisson brackets are
\be
\{P_a, P_b\} = \epsilon_{abc} P^c~, \qquad \{P_a, U_{ij}\} = (U J_a)_{ij}~,
\ee
where we have written the components of $P$ in a basis for the Lie algebra, $P = P_a J^a$. As claimed, $P$ generates $\slt$ operations on $U$ from the right. These brackets can be derived using standard techniques \cite{Faddeev:1988qp}: a quick way to understand them is to note that with this choice of bracket the equations of motion \eqref{eom1} are canonical if we use the Hamiltonian associated with \eqref{ac1}:
\be
\frac{d}{ds}\xi^i = \{H, \xi^i\} \qquad H = - \lam \Tr P^2,
\ee
where $\xi^i$ here denotes either $P$ or $U$.  

We will not carefully quantize this system, referring the reader to \cite{Dzhordzhadze:1994np} for details. We simply state that the resulting Hilbert space is the appropriate highest-weight representation of $\slt \times \slt$. $P_R$ and $P_L$  generate the symmetries of the system. The value of the Casimir on each representation is simply given by
\be
\Tr(P_R^2) = \Tr(P_L^2)  = c_2~.
\ee
Following the notation of \eqref{eq:lwrep}-\eqref{eq:c2rep} we have $\bar h= h$ and
\be
c_2= 2 h(h-1)~.
\ee

We now couple the system to the external gauge fields by promoting \eqref{eq:UG} to a local gauge invariance along the worldline. Recall that the Wilson line follows a path $x^{\mu}(s)$ through an ambient bulk space equipped with connections $(A, \bar{A})$. The bulk is invariant under local gauge symmetries of the form
\be
A_{\mu} \to L(x) \le(A_{\mu} + \p_{\mu}\ri) L^{-1}(x)~, \quad \bA_{\mu} \to R^{-1}(x) \le(\bA_{\mu} + \p_{\mu}\ri) R(x) \ . \label{Agaugesym}
\ee
These gauge transformations have a natural action on the worldline field $U(s)$ through the bulk path $x^{\mu}(s)$. Indeed, if we introduce the following covariant derivative
\be\label{eq:ccdd}
D_s U = \frac{d}{ds} U + A_s U - U \bar{A}_s~, \quad A_s \equiv A_{\mu} \frac{dx^{\mu}}{ds} ~,
\ee
then the global symmetry \eqref{eq:UG} can be promoted to a local gauge symmetry: 
\be
U(s) \to L(x^{\mu}(s))U R(x^{\mu}(s)) \ . \label{Ugaugesym}
\ee
Under such a transformation the covariant derivative transforms homogeneously
\be
{D_sU} \to L(x^{\mu}(s)) ({D_sU} )R(x^{\mu}(s))~,
\ee
and so the following action is invariant under the gauge symmetry given by the combined transformation \eqref{Ugaugesym} and \eqref{Agaugesym}:
\be
S(U,P; {\cal A})_{C} = \int_C ds \le( \Tr\le(P U^{-1} {D_sU}\ri) + \lam(s)\le(\Tr(P^2) - c_2\ri)\ri) \ . \label{gaugedac}
\ee
This action is one of our main results. If the curve $x^{\mu}(s)$ forms a closed loop, then path integration over the fields\footnote{Of course one must also impose the constraint imposed by integrating out $\lam$, though we suppress this dependence for aesthetic reasons.} $U$ and $P$ generates the trace over $\sR$ and computes the Wilson loop, i.e.
\be
W_{\cal R}(C) \equiv {\rm tr}_{\cal R} \left( {\cal P}\exp \oint_C {\cal A}\right) = \int \[{\cal D}U \sD P\]  \exp[-S(U,P; {\cal A})_C] \ .
\ee
If on the other hand we want to evaluate an open-ended Wilson line, then this expression can be viewed as computing a transition amplitude between an initial and final state, i.e
\be
W_{\cal R}(C_{ij}) \equiv \langle j |   {\cal P}\exp \int_{C_{ij}} {\cal A} |i \rangle = \int \[{\cal D}U \sD P\]  \exp[-S(U,P; {\cal A})_{C_{ij}}]~,
\ee
where $C_{ij}$ is an open path and the data specifying $|i \rangle$ and $|j \rangle$ is contained in appropriate boundary conditions on $U(s)$ at the end points of the path. There is a specific choice of boundary conditions that computes entanglement entropy, which is:
\be
U(s = 0) \equiv U_{i} = {\bf 1}~, \qquad U(s = s_f) \equiv U_f = {\bf 1}~, \label{bcU}
\ee
i.e. the identity element of the group at both ends. We will justify this boundary condition in section \ref{sec:cones}.  

In practice we will actually compute this path integral by saddle point, finding a classical solution to $U(s)$ that satisfies the appropriate boundary conditions. We will refer to $U$ as a {\it topological probe}. Note that the dependence on the bulk geometry enters indirectly through the connections $A$, $\bA$ in the gauge-covariant coupling \eqref{eq:ccdd}. A key property of this reformulation is that there is no direct reference to metric-like fields, and we will heavily exploit this feature in the following sections.

\subsection{The geodesic equation} \label{sec:geod}
As we have  argued, this system should be equivalent to that of a massive particle moving in an AdS$_3$ bulk. We now demonstrate one way to see this equivalence and make contact with metric-like fields: in particular we will see that the usual geodesic equation with respect to the metric-like fields makes a somewhat surprising appearance. The background gauge connections $(A, \bA)$ are fixed and determine some bulk geometry. Consider for example computing an open-ended Wilson line denoted by $x^{\mu}(s)$: for convenience we take $s \in [0,s_f]$, and the two endpoints are fixed at $x(s = 0) = x_i$ and $x(s = s_f) = x_f$. Note that as the bulk connections are flat the final answer cannot depend on the actual trajectory taken by the Wilson line (provided it does not wind around a black hole in the bulk), but rather only on its endpoints.   There are also boundary conditions on the probe $U(s)$, which we will discuss along the way.

For the purpose of deriving the geodesic equation it is convenient to eliminate $\lam$ and $P$ from the action \eqref{gaugedac}. Using their classical equations of motion, we find the second order action
\be
S(U;A,\bar A)_C = \sqrt{c_2} \int_C ds \sqrt{\Tr \le(U^{-1}{D_sU}\ri)^2} \ . \label{gaugednewac}
\ee
Note that in this form the action is essentially that of a gauged sigma model.
The equations of motion given by varying  \eqref{gaugednewac}  with respect to $U$ are
\be
\frac{d}{ds}\le((A^u - \bar{A})_{\mu} \frac{dx^{\mu}}{ds}\ri) + [\bar{A}_{\mu}, A_{\nu}^u]\frac{dx^{\mu}}{ds}\frac{dx^{\nu}}{ds} = 0\label{eomfull} \ .
\ee
Here we have made use of the gauge freedom given by reparametrizations of the wordline parameter $s$. In particular we picked $s$ to be the `proper distance' of the probe, i.e the integrand $ \sqrt{\Tr \le(U^{-1}{D_sU}\ri)^2} $ is independent of $s$, which is equivalent to the choice of $\lambda$ being a constant. 

The actual dependence on $U(s)$ in \eqref{eomfull} is in the definition of $A^u$:
\be
A^u_s \equiv U^{-1}\frac{d}{ds} U + U^{-1}A_s U~.
\ee
In these equations $A_{\mu}$ is always contracted with the tangent vector along the path, and so $A_s$ is the only component which matters. 

For reasonable choices of $A, \bA$, these equations of motion are very nontrivial, and their precise form depends strongly on the choice of path $x^{\mu}(s)$. However from the perspective of the equation of motion, $U$ acts like a gauge transformation on the connection $A$. So it seems that a perfectly good ansatz is to look for a solution where the particle does not move in the auxiliary space, i.e. $U(s) = {\bf 1}$; this clearly also satisfies the boundary condition \eqref{bcU}. In this case we find
\be
\frac{d}{ds}\le((A - \bar{A})_{\mu} \frac{dx^{\mu}}{ds}\ri) + [\bar{A}_{\mu}, A_{\nu}]\frac{dx^{\mu}}{ds}\frac{dx^{\nu}}{ds} = 0~. \label{geod2}
\ee
We pause to discuss the interpretation of this equation. It appears to be a differential equation for the {\it path} that the Wilson line takes in the bulk. Of course the choice of path is arbitrary: however this equation tells us that only if the path satisfies this particular differential equation will the condition $U(s) = {\bf 1}$ be a solution to the bulk equations of motion. For a different choice of bulk path $U(s)$ will necessarily vary along the trajectory, but the final on-shell action will be the same. 

As it turns out \eqref{geod2} is actually very familiar. Expressing the connections in terms of the vielbein and spin connection using \eqref{vbspin}, and further using $\om_{\mu}^{\ a} \epsilon_{ab}^{\ \ c} =\om_{\mu\ b}^{\ c}$, we find
\be
\frac{d}{ds}\le(e^{\ a}_{\mu} \frac{dx^{\mu}}{ds}\ri) + \om_{\mu\ b}^{\ a} e^{\ b}_{\nu}\frac{dx^{\mu}}{ds} \frac{dx^{\nu}}{ds} = 0 \ .
\ee
This is precisely the geodesic equation for the curve $x^{\mu}(s)$ on a spacetime with vielbein $e^a$ and spin connection $\om_{\mu\ b}^{\ a}$. It is  equivalent to the more familiar form involving the Christoffel symbols, as can be shown explicitly by relating them to the spin connection and vielbein (see e.g. Appendix J of \cite{Carroll:2004st}). 

Furthermore, on-shell the action \eqref{gaugednewac} for  $U = 1$ reduces to
\bea
S_C &=&  \sqrt{c_2} \int_C ds \sqrt{\Tr \le((A - \bA)_{\mu} (A - \bA)_{\nu} \frac{dx^{\mu}}{ds} \frac{dx^{\nu}}{ds}\ri)}\cr &=& {\sqrt{2c_2}} \int_C ds \sqrt{g_{\mu\nu}(x) \frac{dx^{\mu}}{ds}\frac{dx^{\nu}}{ds}}~, \label{distance}
\eea
which is manifestly the proper distance along the geodesic. Note that the prefactor $\sqrt{c_2}$ indicates that the value of the Casimir controls the bulk mass of the probe, as we alluded to previously.

\begin{figure}[h]
\begin{center}
\includegraphics[scale=0.8]{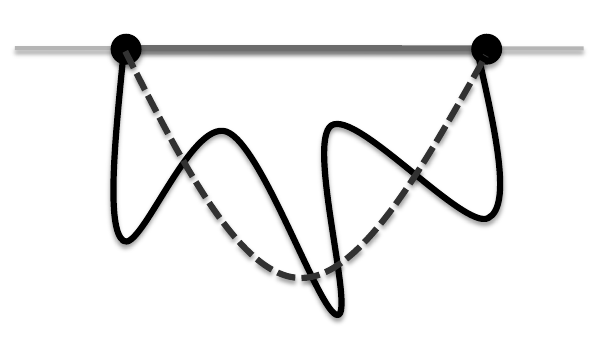}
\end{center}
\vskip -1cm
\caption{The final on-shell action does not depend on the actual bulk path taken, which can be arbitrarily complicated: however the calculation simplifies if it is taken to be a bulk geodesic, as in the dashed line.}
\label{fig:geodpath}
\end{figure}

We have shown that the calculation is simple for a particular choice of bulk path for the Wilson line. However by the flatness of the bulk connections the final result \eqref{distance} must hold for {\it any} path, provided that path can be continuously deformed to a geodesic as illustrated in Figure \ref{fig:geodpath}. {Thus, in the classical limit, we find that the value of the Wilson line between any two points is
\be
W_\sR(x_i, x_f) \sim \exp\le(-\sqrt{2c_2} L(x_i, x_f)\ri),
\ee
where $L(x_i, x_f)$ is the length of the bulk geodesic connecting these two points. Here `$\sim$' denotes the limit $c_2$ large and hence the classical saddle point approximation is valid.}

The somewhat unexpected appearance of the bulk geodesic equation is interesting and (we feel) satisfying: this construction provides a way to obtain geometric data (i.e. a proper distance) from purely topological data (i.e. the flat bulk connections). 


\section{Entanglement entropy revisited}\label{sec:entgrav}

We reviewed in the previous section how for infinite dimensional representations $W_{\cal R}(C)$ can be interpreted as a massive point particle probe of the background solution defined by the flat connection ${\cal A}$, allowing us to compute proper distances along bulk geodesics from purely Chern-Simons data. 

However our ultimate goal is actually to compute {\it entanglement entropies} in the field theory that is dual to this bulk gravity theory. For ordinary Einstein gravity, the holographic entanglement entropy prescription of Ryu and Takayanagi states that these two quantities -- the lengths of bulk geodesics ending on the boundary, and field-theoretical entanglement entropies -- are precisely the same. The recent work \cite{Lewkowycz:2013nqa} has put this statement on a somewhat firmer footing. In this section we will first recast the arguments of \cite{Lewkowycz:2013nqa} in a language appropriate to the Chern-Simons description of gravity. These arguments will generalize to higher spin theories. 

In later parts of this section we further evaluate these Wilson lines using algebraic techniques that rely only on the flatness of the bulk connection. In this way we re-derive standard results from entanglement and thermal entropies in 2d CFTs; again these techniques will also generalize easily to higher spin theories. 

\subsection{Conical singularities and bulk Wilson lines} \label{sec:cones}
We begin by reminding ourselves of the replica trick approach to computing entanglement entropy \cite{CardyCFT,Calabrese:2004eu}. Consider an interval $X$ in a $(1+1)$d quantum field theory in some general state characterized by a density matrix $\rho$. We may construct the {\it reduced} density matrix characterizing degrees of freedom in $X$ by tracing out all degrees of freedom not in $X$, i.e. $\rho_X = \Tr_{\overline{X}} \rho$.
The entanglement entropy of the region $X$ is then the von Neumann entropy associated with $\rho_X$:
\be
S_{\rm EE} = -\Tr(\rho_X \log \rho_X) \ .\label{EEdef}
\ee
The replica trick provides a way to compute this quantity, which we briefly review. We first consider only those $\rho$ which can be obtained from a Euclidean path integral (examples of such states are the vacuum, the finite-temperature state, and states that are obtained from the vacuum by deforming the CFT by an operator such as a chemical potential). Now \eqref{EEdef} can be obtained as follows:
\be
S_{\rm EE} = \lim_{n \to 1} S^{(n)} ~,\qquad S^{(n)} \equiv \frac{1}{1-n} \log \Tr \rho_X^n~,
\ee
where the $S^{(n)}$ are called the $n$-th Renyi entropies. It is easier to compute the Renyi entropies for integer $n$: consider taking $n$ copies of the field theory each defined on a surface with a cut along $X$. We now sew these surfaces together in a cyclic fashion to form an $n$-sheeted surface called $\sR_n$. As is described in detail in e.g. \cite{CardyCFT}, the topology of this surface is such that performing the path integral on $\sR_n$ computes the appropriate traces to evaluate $S^{(n)}$. Analytically continuing the answer to $n \to 1$ we find the entanglement entropy. 

For field theories with gravity duals this is a completely well-posed problem in classical geometry: simply find the bulk AdS$_3$ geometry which asymptotes to the appropriate $\sR_n$ and compute its action. Let us attempt to understand what data characterizes the bulk solution in the limit that $n \to 1$. Take $\sR_n$ and examine one of the endpoints of $X$, denoting it $x_0$. It is clear that we must move around $x_0$ $n$ times to return to the same starting point, and thus the opening angle around $x_0$ is $2\pi n$. Take $\theta$ to be an angular coordinate wrapping around $x_0$; if we take $\theta$ to have periodicity $2\pi$, then the interior geometry as a function of $\theta$ will have an apparent conical deficit of $2\pi(1-\frac{1}{n})$. It is argued in \cite{Lewkowycz:2013nqa} that when the field theory has a gravity dual, this information is enough to usefully characterize the dual bulk geometry and compute its action. Essentially we take the $n \to 1$ limit and extend the conical singularity into the 
bulk: Einstein's equations force this to be done in a unique way that fixes the action of the resulting geometry. 

\begin{figure}[h]
\begin{center}
\includegraphics[scale=0.8]{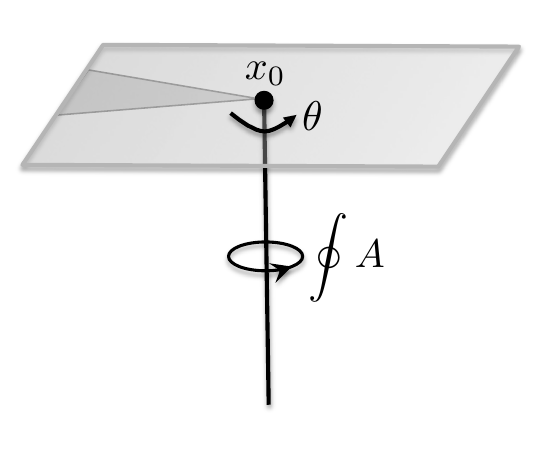}
\end{center}
\vskip -1cm
\caption{Wilson line backreacts on bulk gauge connections, creating a nontrivial holonomy which can be interpreted as a conical singularity.}
\label{fig:backreaction}
\end{figure}

We will now demonstrate those arguments in the Chern-Simons formalism. The required conical singularity in the bulk will be produced by the backreaction of a Wilson line connecting the endpoints of the boundary interval $X$ through the bulk. The strength of the backreaction of the Wilson line is controlled by the quadratic Casimir $c_2$, and we will find a relation between $c_2$ and the deficit angle. To solve for this backreaction, we vary the total Euclidean action
\be
S = iS_{CS}[A] - iS_{CS}[\bA] + S(U;A,\bA)_C
\ee
with respect to $A$ and $\bA$; we find the following equations of motion:
\bea
\frac{ik}{2\pi} F_{\mu\nu}(x) &=&  -\int ds \frac{dx^{\rho}}{ds}\ep_{\mu\nu\rho}\delta^{(3)}(x - x(s)) U^{-1} P U ~,\cr
 \frac{ik}{2\pi} \overline{F}_{\mu\nu}(x)& =& - \int ds \frac{dx^{\rho}}{ds}\ep_{\mu\nu\rho}\delta^{(3)}(x - x(s)) P~. \label{Fsource}
\eea
This states that each Wilson line carries a bundle of $\slt$ flux parametrized by $P$.

Given a specific trajectory for the Wilson line these equations can be explicitly solved. For illustrative purposes we consider pure AdS$_3$ in Poincare coordinates and a Wilson line hanging straight down from the boundary at $x_0$, which we take to be the origin of field theory coordinates, as in Figure \ref{fig:backreaction}. As this trajectory is a bulk geodesic a solution for the Wilson line variables is
\be
\rho(s) = s ~,\qquad U(s) = {\bf 1} ~,\qquad P(s) = \sqrt{2 c_2} J_0 \ . \label{onscon}
\ee
The solution to \eqref{Fsource} that is sourced by this Wilson line, and is asymptotically AdS$_3$ in Poincare coordinates, is given by:
\bea
A & =L a_{\rm source}L^{-1}+LdL^{-1} ~, \quad L = e^{-\rho J_0} e^{-J_{1} z}~,\cr
\bA & =R^{-1} a_{\rm source}R+R^{-1}dR~, \quad R = e^{-J_{-1} \bar z} e^{-\rho J_0}~,
\eea
where the gauge transforms $L$, $R$ generate the asymptotics, whereas the coupling to the source is taken into account by
\be
a_{\rm source}=  \sqrt{\frac{c_2}{2}}\frac{1}{k}\le(\frac{dz}{z} - \frac{d\bar{z}}{\bar{z}}\ri) J_0~.
\ee
 With the help of the identities $\p\le(\frac{1}{\bar{z}}\ri) = \bar{\p} \le(\frac{1}{z}\ri) = \pi \delta^{(2)}(z,\bar z)$,\footnote{We are using conventions where $z=it_E+\phi$ and $\bar z = -it_E+\phi$, and $ d^2 z \equiv dt_E d\phi$. Hence $\delta^{(2)}(z,\bar z)=\delta(t_E)\delta(\phi)$. It is also helpful to remember that the epsilon tensor in complex coordinates is imaginary, $\ep_{\rho z \bar{z}} = \frac{i}{2}$.} one can verify explicitly that these connections satisfy \eqref{Fsource}, i.e. they are flat everywhere except for a well-defined singularity where the Wilson line sources them. 

Now constructing the metric using \eqref{metsl2} we find
\be
ds^2 = d\rho^2 + e^{2\rho}\le(dr^2 + r^2\le(\frac{\sqrt{2 c_2}}{k}-1\ri)^2 d\th^2\ri) ~,
\ee
where we have switched to polar coordinates $(r,\theta)$ on each constant $\rho$ slice; i.e. $z\equiv re^{i\theta}$. We see that this is precisely the metric for AdS$_3$ in Poincare coordinates with a conical singularity surrounding the Wilson line. Demanding the deficit angle to be $2\pi (1-\frac{1}{n})$ and expanding in $n-1$ we find:
\be
\sqrt{2 c_2} = k(n-1) = \frac{c}{6}(n-1)~, \label{c2nrel}
\ee
where in the last equality we have related the bulk Chern-Simons level to the boundary theory central charge ($c=6k$). This is the desired relation between the Casimir characterizing the representation and the conical deficit. Note that the Casimir must scale like the central charge; from a bulk point of view this is because the central charge controls Newton's constant, and so to create a sizable deficit we require a very heavy probe in the limit that bulk gravity is weak. 

To use this to compute the entanglement entropy, we now evaluate the bulk action and take the $n \to 1$ limit. The portion of this action that depends on the interval $X$ is the action of the Wilson line source itself. As demonstrated above, in the semiclassical limit we always find that this on-shell Wilson line takes the form $W_\sR(C) \sim \exp(S_{\rm on-shell}(C)) \sim \exp\le(-\sqrt{2 c_2}L_C\ri)$, where $L_C$ does not depend on $c_2$. Thus the $n$ dependence in this limit factors out, and the entanglement entropy is
\be
S_{\rm EE} = \lim_{n \to 1} \frac{1}{1-n} \log \Tr \rho_X^n = \lim_{n \to 1} \frac{1}{1-n} \log(W_{\sR}(C)) = -
\lim_{n \to 1}\frac{1}{1-n} \sqrt{2 c_2} L_C = \frac{c}{6} L_C~,
\ee
Operationally, performing these series of steps is equivalent to stating that the entanglement entropy is equal to the on-shell Wilson line action if we make the substitution 
\be
\sqrt{2c_2} \to \frac{c}{6} ~,\label{c2rel1}
\ee
in the final answer. This automatically takes care of the $n$-dependence, and we will simply do this from now on, but it should be kept in mind that the motivation for this procedure is actually the reasoning above. 

We may then simply write the entanglement entropy as
\be
S_{\rm EE} = -\log\le(W_{\sR}(C)\ri) \ ,\label{eesl2}
\ee
together with \eqref{c2rel1}. Finally, recall that in the previous section we have actually shown that $L_C$ is the proper distance separating the two boundary points; using the Brown-Henneaux relation \cite{Brown:1986nw},  $c = \frac{3}{2G_3}$, we see that the entanglement entropy is
\be
S_{\rm EE} = \frac{ L_C}{4G_3}~.
\ee
Thus in the case of ordinary Einstein gravity on AdS$_3$ \eqref{eesl2} is precisely equivalent to the usual Ryu-Takayanagi formula, simply written in terms of the Chern-Simons description of AdS$_3$ gravity. Before moving on, we make a few comments:
\ben 
\item 
In a metric treatment of the theory the appearance of a minimal distance may be attributed to the fact that the required conical singularity is created by a massive bulk worldline. The requirement that bulk stress-energy be conserved in its presence is equivalent to the bulk geodesic equation and forces this trajectory to be a minimal surface. In the Chern-Simons treatment the actual bulk trajectory that the Wilson line takes does not matter: nevertheless demanding that the Wilson line variable $U(s)$ satisfy its own equations of motion appears to enforce an equivalent condition which results in the same answer for the on-shell action. 
\item In the following sections we will use the boundary conditions  $U(0) = U(s_f) = {\bf 1}$. To shed some light on this condition, note from \eqref{Fsource} that if $U(s) \neq {\bf 1}$, then we have 
\be\label{eq:noT}
F - \overline{F} \sim \le(U^{-1} P U - P\ri)\delta^{(2)}(x - x(s)) \ ,
\ee
i.e. this difference of connections does not generically vanish along the Wilson line. In a metric formulation of the theory, the difference between $F$ and ${\overline{F}}$ is {\it torsion}. For the purpose of computing gravitational entropy (and entanglement) we  are demanding that the bulk Wilson line is sourcing only  curvature and not  torsion. There are a couple of ways to ensure this. We could demand that $U(s) = {\bf 1}$ everywhere; for $\slt$ this is compatible with the equations of motion, however we will see that it is too strong (and unnecessary) for future generalizations.\footnote{This condition can be weakened somewhat by requiring that $U$ commutes with $P$.} Instead notice that we do have
\be\label{eq:noTU}
UFU^{-1} - \overline{F} =0 \ .
\ee
In this expression $U$ is acting on $F$ as a gauge transformation. Hence if $U$ can be removed via a trivial gauge transformation of $F$---i.e a gauge transformation which does not affect the background state--- there will be a frame where the probe is not generating torsion. 

For the purpose of computing entanglement entropy, we are seeking boundary conditions that are invariant under local Lorentz rotations at the boundary, which in the parametrization \eqref{Agaugesym} are the subgroup of $\slt \times \slt$ with
\be\label{eq:LRU}
L = R^{-1} \ .
\ee 
Requiring that $U(0) = U(s_f) = {\bf 1}$ at the end points  assures Lorentz invariance at the boundary (actually it is the unique choice). Further, if $U(s)$ acts trivially at the boundary we are guaranteed as well that  $U$  can be completed in the interior as a function of $s$ that won't change the state described by $F$.
\een 

It should already be evident that the form \eqref{eesl2} will easily generalize to higher spin theories of gravity, and we will explore this in detail in later sections. Before doing this, we develop some alternative techniques for evaluating \eqref{eesl2}. 

\subsection{The entanglement entropy of an open interval} \label{sec:openI}

In this section we calculate the entanglement entropy of an open interval, i.e. we compute $\sW_{\sR}(C)$ with $C$ ending on the AdS boundary at two points defining an interval of length $L$. As described in Section \ref{sec:geod}, one way to do this is to take $C$ to follow a bulk geodesic, in which case we obtain the proper length. In this section we will present an alternative route to this result that follows from using the $\slt$ gauge invariance. 

We will work in this section with the action in the first-order formulation \eqref{gaugedac}
\be
S(U,P; A,\bA)_{C} = \int ds \le( \Tr\le(P U^{-1} {D_sU}\ri) + \lam(s)\le(\Tr(P^2) - c_2\ri)\ri) \ , \label{gaugedac2}
\ee
where the covariant derivative $D_s$ is given by \eqref{eq:ccdd}. The equations of motion reduce to
\be
U^{-1} D_sU +2\lambda P=0~,\quad  \frac{d}{ds} P  + [\bar{A}_s,P]=0~ \label{eom1st}
\ee
in addition to the constraints $\Tr(P^2)=c_2$. 
It is straightforward to show that these equations are equivalent to the second-order formulation \eqref{eomfull}. On-shell, the action can easily be computed by acting with $P$ on the the left-hand side of the first equation of \eqref{eom1st} and taking a trace:
\bea\label{eq:onact}
S_{\rm on-shell} &=& \int_C ds\Tr\le(P U^{-1} D_sU\ri)\cr
&=&- 2c_2\int_C ds \lambda(s)~.
\eea
Thus to determine the on-shell action, we need to compute the on-shell value of $\lambda(s)$. 

We will compute this in various bulk spacetimes with different connections $A, \bA$. We first consider empty AdS$_3$. The connection describing AdS$_3$ in Poincare coordinates is
\be
A = e^{\rho} J_1 dx^+ + J_0 d\rho~, \qquad \bA =- e^{\rho}J_{-1} dx^- - J_0 d\rho~, \label{ads3conn2}
\ee
where $x^\pm =t\pm \phi$ and $\rho$ is the radial direction. We would like to consider the Wilson line with the following boundary conditions on the spacetime coordinates:
\be
\rho(s=s_f) = \rho(s=0)\equiv  \rho_0~,  \qquad \phi(s=s_f)-\phi(s=0)\equiv \Delta\phi~, \label{path}
\ee
while $t$ is fixed. Here $s$ is the parameter along the path, varying from $s = 0$ to $s = s_f$. This is all we need to specify about the curve; it will be clear from the construction that it is irrelevant if the path $(x^\pm(s), \rho(s))$ is a geodesic or not. 
 
There are probably many ways to construct this solution---one route was already outlined in section \ref{sec:geod}. Here we will take a different route;  we will solve the system with a trick which will exploit the topological nature of the system. In general the difficulty in solving \eqref{eom1st} comes from the fact that $A, \bA$ are nontrivial. Thus consider first solving  the problem in an empty gauge, i.e. the solution in an unphysical ``nothingness'' spacetime with $A = \bar{A} = 0$.  The solution to this problem is immediate: denoting it by $U_0(s)$ and  $P_0$, we have
\be
U_0(s) = u_0 \exp\le(-2 \alpha(s) P_0\ri)~,\quad {d\alpha\over ds} =\lambda~, \label{A0sol}
\ee
with $P_0$  and $u_0$  constant elements and  $\Tr(P_0^2) = c_2$. Thus all solutions are labeled by an element of the group, the starting point $u_0$, and an element of the algebra, the momentum $P_0$. 

However the bulk equations of motion guarantee that the bulk connections are flat: thus every solution is {\it locally} a gauge transform of the ``nothingness'' solution $A = 0$. For the case of Poincare AdS$_3$ we have
\be
A = L d L^{-1}~, \quad L = e^{-\rho J_0} e^{-J_{1} x^+}~, \qquad \bar{A} = R^{-1} d R~, \quad R = e^{-J_{-1} x^{-}} e^{-\rho J_0}~. \label{pgauge}
\ee
To find a solution to the equations on an AdS$_3$ background we can simply take the appropriate gauge transform of the ``nothingness'' solution. A solution to \eqref{eom1st} with connections \eqref{pgauge} is related to the ``nothingness''   in \eqref{A0sol} via
\be
U(s) = L(x(s))U_0(s)R(x(s))~, \qquad P(s) = R^{-1}(x(s)) P_0 R(x(s))~,
\ee
where $x(s)$ is understood to be the path of the Wilson line. This constitutes a general solution to the system.  Further, with this parametrization the on-shell action \eqref{eq:onact} becomes
\bea\label{eq:on2}
S_{\rm on-shell} 
= -2c_2\int_0^{s_f} ds \lambda(s) = -2c_2 \Delta \alpha~,
\eea
where $\Delta \alpha\equiv  \alpha(s_f)-\alpha(0)$. 
All  we need to do is correctly choose $u_0$ and $P_0$ to satisfy the boundary conditions \eqref{path}, which as well will constrain the boundary values of $\alpha(s)$. This makes evident that only the topology of the curve $x^\mu(s)$ is relevant, the path is not necessarily a geodesic.  

Next, we need to specify boundary conditions on the field $U(s)$ and the natural choice is to impose Dirichlet boundary conditions on the interval $[0,s_f]$.  Evaluating $U(s = 0)\equiv U_i$ we find
\be
e^{-\rho_0 J_0}e^{-\phi(0)J_1 }u_0e^{-2\alpha(0)P_0}e^{\phi(0)J_{-1} } e^{-\rho_0 J_0} =U_i ~,\label{eq:u0}
\ee
and for  $U(s = s_f)\equiv U_f$ we have
\be\label{eq:u02}
e^{-\rho_0 J_0}e^{-\phi(s_f)J_1 }u_0e^{-2\alpha(s_f)P_0} e^{\phi(s_f)J_{-1} } e^{-\rho_0 J_0}= U_f ~.
\ee
Solving for $u_0$ in  \eqref{eq:u0} and replacing it in \eqref{eq:u02} gives
\be
\exp\le(-2 \Delta \alpha P_0\ri) =  e^{\phi(0)J_{-1}} \le(e^{\rho_0 J_0}U_i e^{\rho_0 J_0}\ri)^{-1}e^{- \Delta \phi J_1}  \le(e^{\rho_0 J_0}U_f e^{\rho_0 J_0}\ri) e^{-\phi(s_f)J_{-1}}\label{p0eq}\ .
\ee
This equation determines $\Delta \alpha$ and $P_0$ as a function of the boundary conditions of the curve \eqref{path} and of the probe $U_{i,f}$. As argued for in \eqref{eq:noT}-\eqref{eq:LRU}, we will specialize to the boundary conditions
\be
U_i = U_f = {\bf 1} \ .
\ee

We now need to solve explicitly  for $\Delta \alpha$ and determine the on-shell action. Since equation \eqref{p0eq} is independent of the representation of the generators,  the simplest way to extract $\Delta \alpha$ is by choosing a matrix representation and  taking the trace of \eqref{p0eq}. For sake of simplicity, we use the fundamental representation of $SL(2,\RR)$; this gives
\be
{\rm tr}_{f}\exp\le(2 \Delta \alpha P_0\ri) = {\rm tr}_{f}\left[ e^{-2\rho_0 J_0} e^{- \Delta \phi J_1}  e^{2\rho_0 J_0} e^{\Delta\phi J_{-1}}\ri]\label{p0eq2}\ .
\ee  
Note that we have $\tr_{f} P_0^2 = c_2$ and $\tr_f P_0 = 0$, implying that the eigenvalues of $P_0$ in the fundamental representation are $\pm \sqrt{\frac{c_2}{2}}$ and thus that the trace of the left-hand side is
\be
{\rm tr}_f(\exp(2\Delta\alpha P_0)) = 2 \cosh\le( \Delta\alpha \sqrt{2 c_2}\ri)~.
\ee
The trace of the right-hand side may be computed explicitly and is $2 + e^{2\rho_0} (\Delta \phi)^2$. Equating these expressions we find an expression for $\Delta\alpha$
\be
\Delta\alpha =- \frac{\cosh^{-1}\le(1 + \frac{e^{2\rho_0} (\Delta \phi)^2}{2}\ri)}{ \sqrt{2 c_2}} \ .
\ee
We emphasize that this expression for $\Delta\alpha$ is independent of the representation; for instance it is straight forward to check that in the adjoint representation one obtains the same answer as a function of $h$. Finally, evaluating \eqref{eq:on2} gives
\be
S_{\rm on-shell} =  \sqrt{2c_2} \cosh^{-1}\le(1 + \frac{e^{2\rho_0}  (\Delta \phi)^2}{2}\ri) \sim 2\sqrt{2 c_2} \log (e^{\rho_0}  \Delta \phi)~,
\ee
where in the last inequality we have assumed that $e^{\rho_0}\Delta\phi \gg 1$. Recall that in the parametrization of the path \eqref{path} $\Delta \phi$ directly measures the length of the interval, and thus we are assuming that the length of the interval is large in units of the UV cutoff $\epsilon \equiv e^{-\rho_0}$. Further making the substitution $ \sqrt{\frac{c_2}{2}} = \frac{c}{12}$ from \eqref{c2rel1} we find
\be
S_{\rm EE} = \frac{c}{3} \log\le(\frac{\Delta\phi}{\epsilon}\ri)~.
\ee
This is of course the celebrated result from CFT$_2$ and is also the same answer that one finds from solving the bulk geodesic equation. However, this construction does not require the solution of any differential equations and follows from purely algebraic operations.  

Note that the key fact used in this derivation is simply that the bulk connections are flat. Hence the method allows immediate generalization to any solution of AdS$_3$ gravity. Such solutions can be parametrized as
\be
A = b^{-1} (a + d) b~, \qquad \bA = b (\ba + d) b^{-1}~, \qquad b \equiv \exp(\rho J_0) \ .
\ee
Here $a, \ba$ are flat connections with components in the $(t,\phi)$ directions and carry the information of the charges (e.g. mass, angular momentum) of the solution.  The gauge transformation parameter $b$ introduces the radial dependence. For example for the BTZ black hole we have
\be\label{eq:btz} a = \le(J_1 - \frac{2 \pi \sL}{k} J_{-1}\ri) dx^+ ~,\qquad \bar{a} = -\le(J_{-1} - \frac{2\pi\bar \sL}{k}  J_1\ri)dx^-~,
\ee
where $\sL$ and $\bar{\sL}$ are the left and right-moving zero modes of the stress tensor. Following \cite{Gutperle:2011kf}, we are normalizing the modes using the Chern-Simons level $k$ \eqref{eq:level}. In terms of the mass and angular momentum of the black hole we have
\be
\sL={1\over 4\pi}({M -J})~,\qquad \bar \sL={1\over 4\pi}({M +J})~.
\ee

For a general $a, \ba$ the generalization of \eqref{pgauge} is
\bea
R(x^{\pm},\rho) &=& \exp\le( \int_{x_0}^{x} dx^i \ba_i \ri) \exp\le(-\rho J_0\ri)~, \cr L(x^{\pm},\rho)& =& \exp\le(-\rho J_0\ri) \exp\le(-\int^{x}_{x_0} dx^i a_i \ri)~. \label{gengauge}
\eea
Here $i$ runs over the field theory directions, and the integration in the exponents can be taken along any path connecting an arbitrary reference point $x_0$ to the point where we evaluate the gauge transform: as the equations of motion require $a$ and $\ba$ to be flat the path taken is not important. We note that if the boundary has a nontrivial topology (e.g. if the CFT is defined on a cylinder) then generally $L$ and $R$ will not be single-valued around the nontrivial cycles: this is of course equivalent to the statement that the holonomies of the bulk connection can be nontrivial. This does not affect our current computation of the entanglement entropy of a single open interval. 

The generalization of \eqref{p0eq} for the background \eqref{gengauge} is simply 
\be
\exp\le(-2 \Delta \alpha P_0\ri) = \le(R(0)U_i^{-1}L(0)\ri)\le( R(s_f) U_f^{-1} L(s_f)\ri)^{-1} \ .\label{rhseq}
\ee
 We can apply this to the background connection \eqref{eq:btz}. Following precisely the same steps as above (i.e. evaluating the trace of both sides of \eqref{rhseq}) we find, at large $\rho_0$,
\be
2 \cosh\le(\sqrt{2 c_2} \Delta\alpha\ri) \sim \frac{e^{2\rho_0} k}{2\pi \sqrt{\sL\bar{\sL}}} \sinh\le(\sqrt{\frac{2\pi \sL}{k}} \Delta\phi\ri)\sinh\le(\sqrt{\frac{2\pi \bar{\sL}}{k}} \Delta\phi \ri)~.
\ee
This results in an entanglement entropy of 
\be
S_{\rm EE} = \frac{c}{6} \log\le(\frac{k}{2\pi\sqrt{\sL \bar{\sL}}} \frac{1}{\epsilon^2} \sinh\le(\sqrt{\frac{2\pi \sL}{k}} \Delta\phi\ri)\sinh\le(\sqrt{\frac{2\pi \bar{\sL}}{k}} \Delta\phi\ri)\ri)~,
\ee
where as before we have identified $e^{\rho_0}$ with the UV cutoff $\epsilon^{-1}$. This expression corresponds to an entanglement entropy in a CFT in a thermal state with different values of $\sL$ and $\bar{\sL}$ and so with unequal left and right moving temperatures. This expression was previously derived in a holographic context in \cite{Hubeny:2007xt}. If these temperatures are set equal, i.e. ${\cal L}=\bar {\cal L}$, then using the expression for the temperature of the BTZ black hole  $\beta = \pi\sqrt{\frac{k}{2\pi \sL}}$ we find the familiar CFT answer \cite{Holzhey:1994we,CardyCFT}:
\be
S_{\rm EE} = \frac{c}{3} \log\le(\frac{\beta}{\pi \epsilon}\sinh\le(\frac{\pi \Delta\phi}{\beta}\ri)\ri) \ .
\ee

\subsection{Loops and thermal entropy}\label{sec:thgrav}

It is also interesting to consider the case of closed curves, i.e. paths of the form $x^{\mu}(s_f) = x^{\mu}(0)$. There is a intuitive interpretation of the Wilson loop in this case: as a flat connection ${\cal A}= g^{-1}dg$ is transported around a loop, the operator \eqref{eq:WRC} measures whether $g$ is a single valued function or not. In the Chern-Simons language, these are the holonomies of the connection and they uniquely characterize gauge inequivalent classical solutions. Depending on the topology of the loop $W_{\cal R}(C)$ can be given a more geometrical interpretation; e.g. if the loop is a non-contractible cycle, then $W_{\cal R}(C)$ can be thought as quantifying the size of the cycle.  For loops in the infinite dimensional representations $W_{\cal R}(C)$ computes the proper distance around the horizon, which is of course also the thermal entropy.

The topologies we will consider here are of the form $\RR \times D^2$, with  $\RR$ the time direction. The $S^1$ in the disk is described by $\phi\sim \phi +2\pi$ and can be either contractible or non-contractible. Our Wilson loop will be evaluated along the $S^1$ cycle. In contrast to the open interval case, and in accordance to the topology of the loop, for a closed path the probe should be smooth and hence periodic, i.e.
\be
U(s_f) = U(0)~, \qquad P(s_f) = P(0)~. \label{eq:bcbtz}
\ee

The construction of the solution to the system with these boundary conditions will again make use of the ``nothingness'' trick we used in the previous section. We start by taking  $A = \bA = 0$;  the reference solution is 
\be
U_0(s) = u_0 \exp\le(-2 \al(s) P_0 \ri)~, \qquad \frac{d\al}{ds} = \lam(s)~. \label{not3}
\ee 
Again $u_0$ and $P_0$ are constants which characterize the initial conditions of the probe. And as before we can construct the desired solution via a gauge transformation 
\begin{align}
A  = L dL^{-1}~, \qquad \bA = R^{-1} dR~,
\end{align}
with $L(s)$ and $R(s)$ are given by \eqref{gengauge} with the boundary conditions $x^{\mu}(s_f) = x^{\mu}(0)$. 
After the gauge transformation we have
\be
U(s) = L(s) U_0(s) R(s)~, \qquad P(s) = R^{-1}(s) P_0 R(s)~,
\ee
where $L(s)$ and $R(s)$ are evaluated along the path $x^\mu(s)$ of the Wilson loop. 

The boundary conditions \eqref{eq:bcbtz} on $U(s)$ imply that
\be
\exp\le(-2 \Delta \al  P_0 \ri)  = u_0^{-1} \le(L^{-1}(s_f) L(0)\ri) u_0 \le(R(0) R^{-1}(s_f)\ri)~, \label{eq:cons2}
\ee
which we view as an equation for $\Delta\al$. Notice that
\be
R(0)R^{-1}(s_f) = \exp\le(-\int d\phi\;\ba_{\phi}\ri)~, \qquad L^{-1}(s_f)L(0) = \exp\le(\int d\phi\;a_{\phi}\ri)~, \label{eq:holg}
\ee
which are precisely the holonomies of the connection. 
Using \eqref{eq:holg}, we re-write  \eqref{eq:cons2} as
\be
\exp\le(-2 \Delta \al  P_0 \ri)  = u_0^{-1} \exp\le(2\pi a_{\phi}\ri) u_0  \exp\le(-2\pi\ba_{\phi}\ri)~.\label{eq:redthg}
\ee
Here we limited the discussion to cases where $a_\phi$ and $\ba_\phi$ are constant along the path, and have simply performed the integral over $\phi$. 

Demanding the periodicity of $P(s)$, we find
\be
[P_0,R(s_f)R^{-1}(0) ] =0~. 
\ee
This allows us to diagonalize $P_0$ and $\ba_{\phi}$ simultaneously. If we denote by $V$ the matrix that diagonalizes them, then 
\eqref{eq:redthg} reduces to
\be
\exp\le(-2 \Delta \al  \lambda_P \ri)  = ({u_0V})^{-1} \exp\le(2\pi a_{\phi}\ri) u_0 V  \exp\le(-2\pi \bar\lambda_{\phi}\ri) \label{u0V}
\ee
where $\lambda_P$ and $\bar\lambda_{\phi}$ are  diagonal matrices whose entries are the eigenvalues of $P_0$ and $\bar{a}_{\phi}$ respectively. The left-hand side of this equation is a diagonal matrix. Consistency with the right-hand side requires to choose $u_0$ such that $u_0V$ diagonalizes $a_\phi$. With this choice of $u_0$ we find
\be\label{eq:ALC}
-2\Delta\al \lam_{{P}} = 2\pi(\lam_{\phi} - \bar{\lam}_{\phi})~.
\ee
There are several ways to extract from here $\Delta\al$. The simplest is to pick a representation; using the fundamental representation we find 
\be\label{eq:LP}
{\tr}_f (\lam_{{P}}J_0)= \sqrt{c_2\over2}~,
\ee
hence contracting  \eqref{eq:ALC} with $J_0$ and using \eqref{eq:LP}, we find that the on-shell action \eqref{eq:on2} gives
\bea
 -\log W_{\cal R}(C)
= 2\pi \sqrt{{2c_2}} {\tr}_f((\lam_{\phi} - \bar{\lam}_{\phi})J_0) 
\label{entgrav}
\eea
If we evaluate this formula for the BTZ solution \eqref{eq:btz} we find 
\be
 S_{\rm th} =  -\log W_R(C)= 2\pi \sqrt{2\pi k {\cal L}}+ 2\pi \sqrt{2\pi k \bar {\cal L}}~, 
\ee
where we used 
\be\label{eq:mapc}
\sqrt{c_2\over 2}={c\over 12}={k\over 2}~.
\ee  
 $S_{\rm th}$ is precisely the Bekenstein-Hawking entropy of the BTZ black hole. 
 
 At this stage it seems that this derivation of thermal entropy works for any solution that has a compact spatial cycle: it does not distinguish between a black hole, vacuum AdS in global coordinates, and conical solutions. In other words, we did not need to impose any regularity conditions on the solution so it seems as if we are attributing ``entropy'' to any classical configuration. 
 
Upon further inspection, this is not the case. For global AdS it is simple to see where the above derivations break down. The vacuum solutions are characterized by having trivial holonomies around the $\phi$-cycle \cite{Castro:2011iw}, hence the combinations in \eqref{eq:holg} are exactly equal to unity. This implies that in \eqref{eq:redthg} we have 
 \be
 \exp(-2 \Delta \al  P_0 )_{\rm AdS}={\bf 1}~.
 \ee
 $P_0$ is generically not integral since the Casimir of $P$ is related to the mass of the probe. Therefore the only reasonable solution is to have $\Delta\al=0$, which correctly states that the vacuum (horizonless) solutions do not carry entropy.
  
In contrast if we consider conical defect backgrounds, i.e. solutions that geometrically correspond to having delta function sources at the origin, the situation is different. These solutions are not smooth in Lorentzian signature, and the key feature that characterizes them is that the eigenvalues $\lambda_\phi$ and $\bar \lambda_\phi$ are purely imaginary---their holonomies are elliptic whereas the black hole has parabolic holonomies.  The resulting on-shell action \eqref{entgrav} would be purely imaginary. 

\section{Massive probes  in $\slh$ higher spin gravity} \label{sec:probeSL3}

In the previous sections we explained in detail how to compute holographic entanglement entropies in the $\slt$ Chern-Simons formulation of Einstein gravity on AdS$_3$ in terms of a particular Wilson line. In this section we will show that the Wilson line construction generalizes naturally to theories of higher spin gravity, and hence gives a rather simple way to design a massive probe. This will provide a robust framework to discuss generalizations of the geodesic equation which we will use in the following section to compute entanglement and thermal entropy in these theories. For concreteness we will carry out the explicit computations for the higher spin theory based on $\slh$ Chern-Simons theory. 

\subsection{Brief review of $\slh$ higher spin gravity}

In section \ref{sec:sl2grav} we reviewed the formulation of ordinary AdS$_3$ Einstein gravity as two copies of a $\slt$ Chern-Simons theory. As it is well-known, promoting $\slt$ to  $\slh$ results in a convenient representation of a spin-3 theory of gravity \cite{Blencowe:1988gj,Bergshoeff:1989ns,Henneaux:2010xg,Campoleoni:2010zq}; see \cite{Ammon:2012wc} for a recent review. We will briefly review some relevant aspects here to fix notation. 

The action for spin-3 gravity may be written as
\be
S_{\rm HS} = S_{CS}[A] - S_{CS}[\bA]~,
\ee
where the Chern-Simons form remains
\be
S_{CS}[A] = \frac{k_{\rm cs}}{4\pi} \int \Tr\le(A \wedge dA + \frac{2}{3} A \wedge A \wedge A\ri), \label{cs2}
\ee
except that now $A$ and $\bA$ are valued in the $sl(3,\RR)$ Lie algebra. We are denoting the bulk Chern-Simons level by $k_{\rm cs}$ to distinguish it from the effective $k$ controlling various $sl(2,\RR)$ subalgebras, as we discuss later. `Tr' here denotes a trace using the Killing metric on this algebra, and is equal to the matrix trace in the fundamental representation. See Appendix \ref{app:sl3} for further information about our conventions.

This theory has two AdS vacua, corresponding to the two distinct choices of an $\slt$ subgroup inside $\slh$. The correct interpretation of the bulk degrees of freedom depends on the vacuum that we study. In one of them, denoted the {\it principal embedding}, we pick the three $sl(2,\RR)$ $J_a$ generators from the set of $sl(3,\RR)$ generators $\{L,W\}$ as 
$J_a = L_a$, $a = 0, \pm 1$ (see \eqref{sl3gens} for an explicit parametrization of the $sl(3,\RR)$ generators). With this choice the bulk degrees of freedom can be decomposed into a spin $2$ field (the metric $g_{\mu\nu}$ ) and a spin $3$ field $\phi_{\mu\nu\rho}$, defined as
\be\label{eq:metricfield}
g_{\mu\nu} = \ha {\rm \tr}_f \le(e_{\mu} e_{\nu}\ri)~, \qquad \phi_{\mu\nu\rho} = \frac{1}{3!} {\rm \tr}_f \le(e_{(\mu}e_{\nu}e_{\rho)}\ri) \ .
\ee
The equations of motion following from \eqref{cs2} can now be interpreted as describing Einstein gravity on AdS$_3$ interacting nonlinearly with a nontrivial spin-$3$ field \cite{Campoleoni:2010zq}. Under suitable boundary conditions, the classical phase space of solution is described by  ${\cal W}_3$ algebra \cite{Henneaux:2010xg,Campoleoni:2010zq}. In our conventions the central charge of the dual ${\cal W}_3$ CFT is related to the bulk Chern-Simons level as
\be\label{eq:cp}
c_{P} = 24 k_{\rm cs} \ .
\ee

The other vacuum, called the {\it diagonal embedding}, corresponds to the choice of $sl(2,\RR)$ generators $J_0 = L_0/{2}, J_{\pm 1} = \pm W_{\pm 2}/4$. One can verify that these $J_a$'s still obey the $sl(2,\RR)$ commutation relations. Under this choice the bulk degrees of freedom can be interpreted as a  spin-$2$ field, a pair of spin-$1$ $U(1)$ gauge-fields, and a pair of spin-$3/2$ bosonic bulk fields \cite{Ammon:2011nk, Castro:2011fm}. In this case the classical phase space is described by the ${\cal W}_3^2$ algebra \cite{Ammon:2011nk,Campoleoni:2011hg}. The central charge of the corresponding dual  ${\cal W}_3^2$ CFT has a different relation to the bulk Chern-Simons level,
\be\label{eq:cd}
c_{D} = 6 k_{\rm cs}.
\ee
A further important point is that the symmetry algebra of excitations about this vacuum contains a $U(1)$ Kac-Moody algebra with a negative level, and thus the theory is not unitary \cite{Castro:2012bc}. 

There are several more properties for each $sl(2,\RR)$ embedding. We will introduce more features of these two vacua as we require them. 

\subsection{Highest-weight Wilson lines in $\slh$ gravity}
We now turn to the construction of a bulk probe designed to compute entanglement entropy. The conventional notion of a ``proper distance'' is not gauge invariant. However, the Wilson line construction of the previous section has a natural generalization to higher spin theory. Thus, we propose the following: {\it a Wilson line in an appropriate highest-weight representation of $\slh$ measures entanglement entropy in the dual field theory}.  

To compute the Wilson line in such a representation, we simply generalize the probe action \eqref{gaugedac} studied previously:
\be
S(U,P; A,\bA)_C = \int ds \le( \Tr(P U^{-1}D_{s} U) + \lam_2 (\Tr(P^2) - c_2) + \lam_3 (\Tr(P^3) - c_3)\ri)~. \label{sl3probe}
\ee
Now $U$ is an element of the group $\slh$; similarly $P$ is an element of the Lie algebra of $sl(3,\RR)$. We are using the short hand notation 
\be
\Tr(P^2)\equiv  P^aP^b\delta_{ab}~,\quad \Tr(P^3)\equiv  P^aP^bP^c h_{abc}~,
\ee
with $P=P^aT_a$ and $T_a\in sl(3,\RR)$.
The tensors $\delta_{ab}$ and $h_{abc}$ are fully symmetric Killing forms of the algebra which define the quadratic and cubic Casimirs; see appendix \ref{app:sl3} for detailed definitions. The action \eqref{sl3probe} clearly has a local gauge invariance under two copies of $\slh$, where the action of the global symmetries on the fields is given by 
\be\label{eq:UG3}
U(s)\to L U(s)R ~,\quad  P \to R^{-1} P(s) R ~, \qquad {L,R}\in SL(3,\RR)~.
\ee
Whereas previously we had to constrain only a single Casimir $c_2$, we now need to constrain both a quadratic and a cubic Casimir, and thus we have two Lagrange multipliers: $\lam_2$ and $\lam_3$. Just as before, we claim that a path integral over the field $U$ generates a trace over the appropriate representation. We also need to specify boundary conditions on $U$ at the ends of the open interval: just as in the $\slt$ case \eqref{bcU} we choose
\be
U(s = 0) = {\bf 1}~, \qquad U(s = s_f) = {\bf 1} \ .
\ee
In the higher spin case we have somewhat less justification for this choice, and it should probably be viewed as an assumption. 

 We now briefly discuss properties of a highest-weight representation of $\slh$. Just as in the $\slt$ case, we begin by considering a highest-weight state $|h, w\rangle$ with definite eigenvalues under the elements of the $\slh$ Cartan $L_0, W_0$:
\be
L_0 |h, w \rangle = h |h, w\rangle ~,\qquad W_0 |h, w\rangle = w |h,w \rangle~,
\ee
and which is annihilated by the positive modes of the algebra:
\be
L_1 |h,w\rangle = 0 ~,\qquad W_{1,2} |h, w \rangle = 0 ~.
\ee
We may now generate other excited states by acting with $L_{-1}, W_{-1,-2}$ on this ground state, filling out an {irreducible representation. }

The relationship between  $(h,w)$ and the quadratic and cubic Casimirs is as follows. Starting from the definition \eqref{eq:c2c3}, we have
\be
C_2 = \ha L_0^2 + \frac{3}{8} W_0^2 + \cdots~, \qquad C_3 = {3\over 8}W_0\le( L_0^2 - \frac{1}{4} W_0^2\ri) + \cdots~.
\ee
Here we have omitted all terms which contain the raising and lowering operators $L_{\pm 1}, W_{\pm 1, 2}$; in a fully quantum theory these must be normal-ordered to annihilate the highest-weight state. Acting with $C_2$ and $C_3$ on the highest weight state we find
\be
c_2 = \ha h^2 + \frac{3}{8} w^2~, \qquad c_3 = {3\over 8}w\le( h^2 - \frac{1}{4} w^2\ri)~. \label{cascharges}
\ee
The omitted terms above result in corrections that are subleading in the large $h$, $w$ limit. 
Thus, to compute a trace in a highest-weight representation with charges $(h,w)$, we simply find a solution to the classical equations of motion of the probe action \eqref{sl3probe}, with $c_2$, $c_3$ chosen according to \eqref{cascharges}.

We now turn to the question of what these charges should be for the probe appropriate to the entanglement entropy in a higher spin theory. To understand this, we revisit the conical singularity arguments of Section \ref{sec:cones}. Recall that in ordinary $\slt$ gravity the Wilson line created a conical singularity in the bulk metric. We now seek to create the same singularity in the higher spin case. This requires $w = 0$ and thus $c_3 = 0$: a nonzero $c_3$ will source the higher spin fields. To determine the remaining relation between $c_2$ and the boundary theory central charge, we repeat precisely the same arguments as in Section \ref{sec:cones}, i.e. we compute the backreaction of the Wilson line on the geometry. Just as in \eqref{onscon}, we pick a probe:
\be
\rho(s) = s~, \qquad U(s) = {\bf 1}~, \qquad P(s) = \sqrt{\frac{c_2}{2}} L_0 \ . \label{eq:3onscon}
\ee
and solve for its backreaction, constructing a solution which has a conical singularity but asymptotes to either the principal or diagonal embedding AdS$_3$ vacuum. 

Direct computation shows that even for a fixed $c_2$ and $k_{\rm cs}$ the strength of the effective conical singularity created is {\it different} in the two vacuua: this is because of the different normalization of the generators of the effective $\slt$ in each case. The resulting relations (i.e. the analog of \eqref{c2rel1}) are for the principal and diagonal embeddings respectively:
\be
c_{2P} \to \ha\le(\frac{c_P}{12}\ri)^2 ~,\qquad c_{2D} \to 2\le(\frac{c_D}{12}\ri)^2 ~,\qquad c_{3P,D} = 0 \label{casPD}
\ee 
where we have also used \eqref{eq:cp} and \eqref{eq:cd} to relate the bulk Chern-Simons level with the boundary theory central charges. 

We may also write these relations in terms of $(h,w)$. We have used $h$ to denote the eigenvalue of the operator $L_0$ in the $\slh$ algebra. This is not necessarily the same as the conformal dimension under the $\slt$ subalgebra since there may be relative factors relating the relevant $\slt$ operator ($J_0$ in our notation) to $L_0$: the principal embedding has $L_0 = J_0$, but the diagonal embedding has $L_0 = 2 J_0$. Denoting the conformal dimensions under the respective $\slt$'s as $h_{P,D}$, the charge assignments above can be written:
\be\label{eq:qnsl3}
h_{D,P} \to {c_{D,P}\over 12}~, \quad w=0~.
\ee

Note that while $w = 0$ implies $c_3 = 0$, the reverse is not true, and thus \eqref{eq:qnsl3} is a more complete specification of the charges than \eqref{casPD}. This ambiguity will play a role in our later analysis. 

We believe that a probe satisfying \eqref{eq:qnsl3} is the object closest to being a higher-spin gauge-invariant generalization of the notion of a ``proper distance''.

\subsection{Equations of motion and the lack of geodesics}\label{sec:lack}
The first-order equations of motion following from \eqref{sl3probe} are the generalization of \eqref{eom1st}:
\begin{align}
U^{-1} {D_sU} + 2\lam_2 P + 3\lam_3 (P \times P)  = 0~, \qquad D_sP = 0 , \label{first3}
\end{align}
together with the constraints $\Tr(P^2) = c_2, \Tr(P^3) = c_3$. We define a product $P \times P$ in the $sl(3,\mathbb{R})$ algebra as 
\be
P \times P \equiv h_{abc}T^aP^bP^c~,
\ee
and just as in the lower spin case the covariant derivatives are defined as 
\be
D_sU = \frac{dU}{ds} + A_s U - U \bA_s~, \qquad D_sP = \frac{dP}{ds} + [\bA_s, P]~.
\ee

To compute an entanglement entropy in a higher spin theory we will follow precisely the same prescription as previously: we fix two points $x_0$ and $x_f$ on the boundary and consider a path in the bulk $x^{\mu}(s)$ connecting these two points. We then seek a solution to the bulk equations of motion such that $U(s = 0) = U_i$ and $U(s = s_f) = U_f$. The entanglement entropy will be given by the value of the on-shell action, which is again easily found by multiplying \eqref{first3} with $P$ and taking a trace:
\be\label{eq:3act}
S_{\rm on-shell} = \int_C ds \Tr\le(P U^{-1} D_sU\ri) = \int_C ds \le(-2 \lam_2(s) c_2\ri),
\ee
where we have used $c_3 = 0$. Thus the goal is again to determine the on-shell value of the Lagrange multiplier $\lam_{2}(s)$.

There is actually an important difference between these higher spin equations and those for ordinary $\slt$ gravity. Recall that in Section \ref{sec:geod} we demonstrated that the $\slt$ equations of motion could be trivially solved by taking the path to $x^{\mu}(s)$ to lie on a bulk geodesic. Let us attempt to repeat those steps here.

 Due to the appearance of a cube of the momentum $P$ in the action \eqref{sl3probe}, it is difficult to cast the action itself in a second-order form. Nevertheless, if we pick a gauge where $\lam_2$ and $\lam_3$ are constant in $s$,\footnote{Note that in the spin-2 case where we had only one Lagrange multiplier $\lam(s)$ it was clear that it could  always be made constant via a choice of parametrization of the path. In the spin-3 case it is not clear that both independent Lagrange multipliers $\lam_2(s)$ and $\lam_3(s)$ can be made constant simultaneously; however for the illustrative purposes of this section this possibility is not important, and later on we will not assume this.}  we can find second-order equations of motion by taking a gauge-covariant derivative of \eqref{first3} to find
\be
\frac{d}{ds}\le((A^u - \bar{A})_{\mu} \frac{dx^{\mu}}{ds}\ri) + [\bar{A}_{\mu}, A_{\nu}^u]\frac{dx^{\mu}}{ds}\frac{dx^{\nu}}{ds} = 0 ~,\label{eomfull3}
\ee
where $A^{u} = U^{-1} \le(A_s + \frac{d}{ds}\ri)U$, i.e. superficially the same second-order equations of motion as in the lower-spin case \eqref{eomfull}. In the spin-2 case, this is an equation with three independent components, as $A$ and $\bA$ are valued in the $\slt$ algebra. Recall that in the spin-$2$ case, if we fix $U(s) = {\bf 1}$ then this equation could be interpreted as a differential equation -- indeed, the geodesic equation -- for the three components $x^{\mu}(s)$ of the path. The coincidence that $3$ (the dimension of the $\slt$ algebra) $= 3$ (the dimension of the bulk space) was crucial in guaranteeing that a solution to these differential equations could always be found. 

In the higher spin case, this breaks down: now $A$ and $\bA$ are elements of $\slh$ and generically \eqref{eomfull3} has eight independent components. Thus if we fix $U(s) = {\bf 1}$, we end up with eight differential equations constraining three functions $x^{\mu}(s)$. Generically this is an overconstrained system and has no solution; this means that there is no choice of path $x^{\mu}(s)$ for which we can keep $U(s)$ constant along the trajectory. As we will see, this will not be a serious obstacle in terms of solving the system, but in terms of interpretation it does indicate a fundamental new ingredient in higher spin gravity: on a generic higher spin background  the differential equation \eqref{eomfull3} can no longer be interpreted as a geodesic equation. Of course if $A, \bA$ live only in an $\slt$ subgroup of $\slh$ then all but three of the components of \eqref{eomfull3} are identically zero, and the calculation is precisely equivalent to those performed in the previous section.

\section{Thermal entropy from Wilson lines}\label{sec:thermhs}
In the next two sections we will assume the validity of the conjecture above and use it to compute entanglement entropies in various higher spin backgrounds. An important consistency check on the validity of this prescription is that if we consider a closed Wilson loop that encloses a black hole horizon, we should reproduce the {\it thermal} entropy of the black hole. In this section we will explain how to use our prescription to compute thermal entropies and work out two examples, black holes with higher spin charges in the diagonal and principal embeddings. For the diagonal embedding black holes we find agreement with the literature \cite{Castro:2011fm}. However in the case of principal embedding black holes there are at the moment {\it two} inequivalent formulas for the entropy in the literature \cite{Campoleoni:2012hp,Gutperle:2011kf,Perez:2013xi,Perez:2012cf,deBoer:2013gz}; of course we find agreement with only one of them. We comment on this further below. 

\subsection{Closed paths around horizons}
Consider a closed trajectory $x^{\mu}(s)$ that encloses a black hole horizon, i.e.
\be
x^{\mu}(s_f) = x^{\mu}(0) \ .
\ee
Following the discussion in section \ref{sec:thgrav}, we pick the trajectory to be the $S^1$ cycle with periodicity $\phi\sim \phi+2\pi$. As in the gravitational case, the holonomies of the connections around this horizon are nontrivial and carry information regarding the charges that the black hole carries (mass, angular momentum, higher spin charge, etc.). We will show that our Wilson loop correctly extracts from this data an expression for the thermal entropy. 

We seek a solution to the equations of motion \eqref{first3} that satisfies
\be
U(s_f) = U(0)~, \qquad P(s_f) = P(0)~, \label{closedbc}
\ee
and so is continuous around the closed trajectory. To find this solution we follow precisely the same techniques as in \ref{sec:thgrav}, i.e. we use the gauging up from ``nothingness'' trick. Consider then first a reference solution $U_0(s)$, $P_0$ to the equations of motion \eqref{first3} on the ``nothingness'' spacetime with $A = \bA = 0$:
\be
U_0(s) = u_0 \exp\le(-2 \al_2(s) P_0 - 3 \al_3(s)(P_0 \times P_0)\ri)~, \qquad \frac{d\al_i}{ds} = \lam_i(s)~. \label{not3b}
\ee 
where $P_0 \times P_0 =h_{abc}T^aP^a_0P^b_0$. This solution is characterized by $u_0$, an element of the group $\slh$, and $P_0$, an element of the algebra that satisfies $\Tr(P_0^2) = c_2$, $\Tr(P_0^3) = c_3 = 0$. 

Just as in \eqref{gengauge}, the higher spin black holes of interest may be related to the ``nothingness'' solution via the following gauge transformation:
\begin{align}
A & = L dL^{-1}~, \qquad L(x^{\pm}, \rho) = \exp(-\rho L_0) \exp\le(-\int_{x_0}^{x} dx^i a_i\ri) ~,\\
\bA & = R^{-1} dR~, \qquad R(x^{\pm}, \rho) = \exp\le(\int_{x_0}^x dx^i \ba_i\ri) \exp\le(-\rho L_0\ri)~,
\end{align}
where $a_i$ and $\ba_i$ are constant connections carrying the information of the black hole charges and the integral over $x^i$ is taken in the $x^\pm=t\pm\phi$ directions from a suitable starting point $x_0$. It is important to note that these gauge transformations are not single-valued as we move around the horizon; thus after the transformation $A$, $\bA$ will have nontrivial holonomies. 

Under this gauge transformation the nothingness solution \eqref{not3b} is transformed to
\be
U(s) = L(s) U_0(s) R(s)~, \qquad P(s) = R^{-1}(s) P_0 R(s)~,
\ee
where $L(s)$ and $R(s)$ are evaluated along the path $x^\mu(s)$ of the Wilson line. Now imposing the boundary conditions \eqref{closedbc} we find the nontrivial constraints
\be
[P_0,R(s_f)R^{-1}(0) ]=0~,\label{cons1} 
\ee
and
\be
\exp\le(-2 \Delta \al_2 P_0 - 3 \Delta \al_3 (P_0 \times P_0)\ri)  = u_0^{-1} \le(L^{-1}(s_f) L(0)\ri) u_0 \le(R(0) R^{-1}(s_f)\ri)~. \label{cons2}
\ee
Here $\Delta \al_i \equiv \al_i(s_f) - \al_i(0)$. And as before 
\be
R(0)R^{-1}(s_f) = \exp\le(-\int d\phi\;\ba_{\phi}\ri)~, \qquad L^{-1}(s_f)L(0) = \exp\le(\int d\phi\;a_{\phi}\ri)~, \label{hol}
\ee
i.e. precisely the holonomies around the horizon. 

The on-shell action \eqref{eq:3act}is related to $\Delta \al_2$; thus the challenge is to find $\al_2(s)$ (and if necessary $u_0$, $P_0$, $\al_3$) subject to these constraints. This is not a difficult problem, but it is useful to introduce some notation to keep track of the information. 

The simplest way to solve for $\al_2(s)$ is to evaluate the above expression in a matrix representation of the algebra. We emphasize that this is just a short cut valid in the classical limit. We will use the fundamental representation of $\slh$. In this case we have
\be
P_0 \times P_0 = h_{abc}T^aP_0^bP_0^c=P_0^2 - \frac{c_2}{3}{\bf 1}_{3\times 3}~.
\ee
 Let us  define 
\bea
\mathbb{P} & \equiv & -2 \Delta \al_2 P_0 -   3 \Delta \al_3 (P_0 \times P_0) \cr &=&-2 \Delta \al_2 P_0 - 3 \Delta \al_3 \le(P_0^2 - \frac{c_2}{3}{\bf 1}_{3\times 3}\ri) \ ,
\eea
(the last line being valid for the fundamental representation). The characteristic polynomial of any matrix can be worked out in terms of traces of powers of the matrix; see appendix \ref{app:matrix}. Using this fact, we can easily find the eigenvalues of $\mathbb{P}$ in terms of $c_2$:
\be
\lam_{\mathbb{P}} = \mbox{diag}\le(\ha\le(-2 \sqrt{2 c_2} \Delta \al_2 - c_2 \Delta\al_3\ri),c_2 \Delta \al_3, \ha\le(2 \sqrt{2c_2} \Delta \al_2 - c_2 \Delta \al_3\ri) \ri)\ . \label{eigbigP}
\ee
Next, note from \eqref{cons1} that $P_0$ commutes with $\le(R(0) R^{-1}(s_f)\ri)$, and we may thus diagonalize them simultaneously. Let $V$ be the matrix that performs this diagonalization, i.e.
\be
R(0)R^{-1}(s_f) = V \exp\le(-2\pi \bar{\lam}_{\phi}\ri) V^{-1}, 
\ee
where we have used \eqref{hol} and $\bar{\lam}_{\phi}$ is a diagonal matrix whose entries are the eigenvalues of $\bar{a}_{\phi}$. Performing a similar diagonalization of $\exp(\mathbb{P})$, we may write \eqref{cons2} as
\be
\exp\le(\lam_{\mathbb{P}}\ri) = (u_0 V)^{-1} \exp\le(2\pi\;a_{\phi} \ri) u_0 V \exp\le(-2\pi\;\bar{\lam}_{\phi}\ri)~.
\ee

Now note that the left-hand side of this expression is diagonal. Thus to satisfy this equation we must pick $u_0$ to be an $\slh$ matrix such that the product $(u_0 V)^{-1}$ diagonalizes $\exp\le(2\pi\;a_{\phi}\ri)$. Picking such a $u_0$ we find the simple relation
\be\label{eq:matcheigen}
\lam_{\mathbb{P}} = 2\pi(\lam_{\phi} - \bar{\lam}_{\phi})~.
\ee 
Through \eqref{eigbigP} the solutions to this equation determine $\Delta\al_{2,3}$ and thus the on-shell action. Note however that in writing expressions of this form we have made several choices about eigenvalue ordering. Below we pick one particular ordering for the eigenvalues of $\mathbb{P}$ -- that which is given in \eqref{eigbigP} --  and denote it from now on as the {\it primary ordering}. We will return to the implications of the this choice shortly.

Note that given \eqref{eigbigP} and the explicit expression for $L_0$ in \eqref{sl3gens} we may solve for $\Delta\al_2$ 
\be\label{eq:alpha2}
\Delta\al_2 = -\frac{2\pi}{2\sqrt{2 c_2}}{\tr}_f((\lam_{\phi} - \bar{\lam}_{\phi})L_0)~,
\ee
and thus the entropy works out to
\be
S_{\rm th} = 2\pi \sqrt{\frac{c_2}{2}} {\tr}_f((\lam_{\phi} - \bar{\lam}_{\phi})L_0) \ . \label{entexp}
\ee
Precisely the same expression for the entropy of a higher spin black hole was derived previously in \cite{deBoer:2013gz} from a study of the on-shell Euclidean action. This is a nontrivial test of our formalism.  


\subsection{Diagonal embedding black hole}\label{sec:entDE}
We now evaluate this expression for some specific examples of higher spin black holes. Much of this analysis was already performed in \cite{deBoer:2013gz}; we will review some of their results, and would like to take this opportunity to discuss the physical implications of the choice of eigenvalue ordering made in \eqref{eigbigP}. 

We consider first black holes in the diagonal embedding \cite{Castro:2011fm}, where we have
\bea
a &=& (W_2 + \om W_{-2} - q W_0) dx^+ + \frac{\eta}{2} W_0 dx^-~, \cr \ba &=& (W_{-2} + \om W_2 - q W_0) dx^- + \frac{\eta}{2} W_0 dx^+ \ , \label{diagBH}
\eea
where $w$, $q$ and $\eta$ are constants representing the mass, charge and chemical potential of the black hole, respectively. 
In the diagonal embedding the field content includes a metric coupled to a pair of $U(1)$ Chern-Simons gauge fields: this black hole solution may be viewed as a BTZ black hole with nontrivial $U(1)$ holonomies around the horizon; these $U(1)$ holonomies here are the manifestation of the nontrivial $\slh$ structure. Denoting these $U(1)$ gauge fields by $\chi$, $\bchi$ we find \cite{Castro:2011fm}
\be
\chi  = \frac{\eta}{2} dx^- - q dx^+~, \qquad \bchi = \frac{\eta}{2} dx^+ - q dx^-~.
\ee
Horizon regularity can be shown to require that $\eta = 2q$, which is equivalent to demanding that the time component of these gauge fields vanish at the horizon. 

We now evaluate the eigenvalues to find:
\begin{align}
\lam_{\phi}  = \mbox{diag}\le(\frac{1}{3}\le(-2 q - \eta - 12 \sqrt{\om}\ri), \frac{2}{3}(2q + \eta), \frac{1}{3}\le(-2 q + 12 \sqrt{\om} - \eta\ri)\ri)~, \label{lamdiag}
\end{align}
and $\bar{\lam}_{\phi}  = -\lam_{\phi}$. Evaluating \eqref{entexp} and using the relation \eqref{casPD}, we find the entropy to be
\be
S_{\rm th} = 2\pi \frac{c_D}{12} 16 \sqrt{\om}~, \label{diagtherm}
\ee
with $c_D$ the central charge of the theory in the diagonal embedding. Note that the values of the $U(1)$ holonomies have dropped out of the  final answer. Though we have obtained it in a formalism that was manifestly $\slh$ invariant, this entropy is actually equal to the usual Bekenstein-Hawking entropy, i.e. the area of the horizon in the metric representation of the theory. 

As promised, we would now like to discuss the implications of the eigenvalue ordering choice made in \eqref{eigbigP}. There are six possible orderings; however the equations have a symmetry under $\Delta\al_2 \to -\Delta\al_2, P_0 \to -P_0$, so only three of these orderings are distinct and correspond to physically reasonable (i.e. positive) answers. To understand how we choose amongst them, note first that $\Delta\al_{3}$---the Lagrange multiplier constraining the cubic Casimir---is one of the new ingredients in the higher spin theory. If we require that the $\slh$ results smoothly match on to the $\slt$ results as all of the higher spin ingredients are turned off, it is necessary to demand that $\Delta\al_3$ vanishes in this limit. This was the reason for our choice of the ordering \eqref{eigbigP}; note from comparison to \eqref{lamdiag} that as $\eta, q \to 0$, we find that $\Delta\al_3 \to 0$ as well. We denote this choice the {\it primary ordering}. 

Thus if we demand continuity in the $\slt$ limit a single ordering is picked for us. Nevertheless it is instructive to understand the physical significance of the other orderings.  Consider then the different choice:
\be
\lam'_{\mathbb{P}}  = \mbox{diag}\le(c_2 \Delta \al_3, \ha\le(-2 \sqrt{2 c_2} \Delta \al_2 - c_2 \Delta\al_3\ri),\ha\le(2 \sqrt{2c_2} \Delta \al_2 - c_2 \Delta \al_3\ri) \ri) \ . \label{eigbigPnew}
\ee
Repeating the same steps as above, we now find for the ``entropy'':
\be
S' = 4\pi \sqrt{\frac{c_2}{2}} \le(2q + \eta + 4 \sqrt{\om}\ri),
\ee
where the answer depends on $q$ and $\eta$, i.e. on the background $U(1)$ gauge flux threading the horizon. In fact the dependence is as though our probe had a $U(1)$ charge  of $2i\sqrt{\frac{c_2}{2}}$ under the gauge fields $(\chi, -\bchi)$. 

To understand this result note that our construction only constrains the Casimirs of our probe. In particular, by setting $c_3 \to 0$ we attempted to guarantee that all higher spin charges carried by the probe were zero. The equation for the cubic Casimir \eqref{cascharges} is
\be
c_3 = w\le(\frac{3}{8} h^2 - \frac{3}{32} w^2\ri) \label{cascharges2} \ .
\ee
Recall that $w$ is the eigenvalue of the highest-weight state under $W_0$. Note that there are multiple ways to obtain $c_3 = 0$; we may set $w = 0$ (corresponding to no higher spin charges); alternatively we may set $w = \pm 2 h$, corresponding to a nonzero $W_0$ charge of $2\sqrt{\frac{c_2}{2}}$. In the case of the diagonal embedding this maps to a nonzero $U(1)$ charge for the bulk probe: this is precisely what has happened above.\footnote{It is interesting that the $U(1)$ charge appears to be imaginary, as it couples exponentially rather than as a phase. We believe this is related to the non-compactness of the gauge group. The kinetic term for the $U(1)$ field in $SL(3,\RR)$ has the opposite sign; this could be fixed by multiplying the current by a factor of $i$, but at the cost of generating complex charges, which may be what has happened here. This reflects as well on the non-unitarity of the theory: the current dual to the bulk gauge field has a Kac-Moody algebra with negative level \cite{Castro:2012bc}.
 }

Thus in performing our computations care must be taken to guarantee that we are working always with the representation with $w = 0$, i.e. the primary ordering. This can be ensured by making sure that our choices can always be continuously connected to the $\slt$ results.

\subsection{Principal embedding black hole}
The evaluation of  \eqref{entexp} for the case of a black hole in the principal embedding was worked out in detail in \cite{deBoer:2013gz}, and so here we will simply review their results to set the stage for later. The connection for a non-rotating black hole  can be written:
\bea
a & = &\le(L_1 - \frac{2\pi}{k} \sL L_{-1} - \frac{\pi}{2k} \sW  W_{-2}\ri) dx^+\cr &&+ \mu\le(W_2 - \frac{4\pi \sL}{k} W_0 + \frac{4\pi^2 \sL^2}{k^2}W_{-2} + \frac{4\pi \sW}{k}L_{-1}\ri) dx^-~, \cr
\ba & = &-\le(L_{-1} - \frac{2\pi}{k} \sL L_{1} + \frac{\pi}{2k} \sW W_{2}\ri) dx^-\cr &&  + \mu\le(W_{-2} - \frac{4\pi \sL}{k} W_0 + \frac{4\pi^2 \sL^2}{k^2}W_{2} - \frac{4\pi \sW}{k} L_{1}\ri) dx^+~. \label{princBH}
\eea
Here $\mu$ is the chemical potential and $\sW$ the spin-3 charge. $k$ in these expressions can be thought of as the effective bulk Chern-Simons level for the appropriate $\slt$ subgroup of $\slh$ and is related to the central charge by $c_P = 6 k$. Regularity at the horizon enforces relations between them; as shown in \cite{Gutperle:2011kf,Ammon:2011nk} these constraints can be explicitly solved in terms of a dimensionless parameter $C$:
\be
\sW = \frac{4(C-1)}{C^{3/2}} \sL \sqrt{\frac{2\pi \sL}{k}}~, \qquad \mu = \frac{3\sqrt{C}}{4(2C - 3)}\sqrt{\frac{k}{2\pi \sL}}~.
\ee
Note that $C \to \infty$ is the limit in which the higher spin charge vanishes. The eigenvalues may be worked out to be
\be
\lam_{\phi}  = 2\sqrt{\frac{2\pi \sL}{k}}\mbox{diag}\le(\frac{3+C \left(-2+\sqrt{-3+4 C}\right)}{\sqrt{C} (-3+2 C)},\frac{2}{\sqrt{C}},\frac{3-C \left(2+\sqrt{-3+4 C}\right)}{\sqrt{C} (-3+2 C)}\ri) ~,
\ee
and $\bar{\lam}_{\phi}  = -\lam_{\phi}$.
Now evaluating \eqref{entexp} and using \eqref{casPD} as well as $c_P = 6k$ we find
\be
S_{\rm th} = 4\pi\sqrt{2\pi k \sL}\frac{\sqrt{1 - \frac{3}{4C}}}{1 - \frac{3}{2 C}}~. \label{thermprinc}
\ee

As we alluded to before, for the higher spin black hole \eqref{princBH} there are two distinct thermodynamic entropies that can be attributed to the solution. The two entropies are thought to differ due to different notions of ``energy''. The original derivation \cite{Gutperle:2011kf} used a notion of energy following from considerations of the OPE of the boundary theory stress tensor; however some subsequent derivations have used a notion of energy that follows naturally from the bulk gravitational Hamiltonian \cite{Perez:2012cf,Perez:2013xi,deBoer:2013gz}. It was made clear in \cite{deBoer:2013gz} that these notions of energy disagree in the presence of a source $\mu$, and thus lead to different entropies following from the First Law. There is also an independent derivation based on expanding the theory in terms of metric-like fields and using the conventional Wald formula \cite{Campoleoni:2012hp}: this can only be done to lowest order in the higher spin charge, but it makes no explicit reference to a 
boundary theory energy. It agrees with derivations that use the bulk Hamiltonian for the energy \cite{Perez:2012cf,Perez:2013xi,deBoer:2013gz}. 

Our calculation {\it also} agrees with the computations that use the bulk Hamiltonian, and so disagrees with the original calculation \cite{Gutperle:2011kf}. As our computation can be thought of as implementing the conical deficit approach to computing black hole entropy,\footnote{We also note the recent work \cite{Kraus:2013esi} that motivates the original entropy formula \cite{Gutperle:2011kf} using conical singularities; however that computation involves the regulation of a singular action and appears to depend on the manner in which the singularity is regulated. We believe the unambiguous way to regulate that action is to include the source that is creating the singularity, which is what we have done in this work.} this seems consistent with general arguments (from a metric formulation) that the Wald and conical deficit approaches to computing black hole entropy are equivalent \cite{Iyer:1995kg}. It thus appears that implicitly our probe is coupling to the bulk Hamiltonian. It might be possible to design 
a probe that somehow couples instead to the boundary stress tensor by tweaking the probe in a controlled manner. This would allow us to reproduce the entropy in \cite{Gutperle:2011kf}; unfortunately we haven't been able to argue that any of these tweaks are either physical and/or natural.

\section{Higher spin entanglement entropy of an open interval} \label{sec:EEhs}
Having exhaustively discussed thermal entropies, we now turn to the computation of the {\it entanglement} entropy of an open interval. Recall that we seek a solution to the equations of motion \eqref{first3} along a bulk Wilson line with two endpoints at the AdS boundary (i.e. at a large value of $\rho \equiv \rho_0$) separated in the $\phi$ direction by a distance $\Delta\phi$. We take the boundary conditions on this solution to be
\be
U(s = 0) = U_i ~,\qquad U(s = s_f) = U_f \ . \label{bcsl3}
\ee
We will follow precisely the same techniques used in the $\slt$ case in Section \ref{sec:openI}. The larger dimension of the matrices involved makes the problem slightly more involved operationally, but the strategy remains the same.  We start from the ``nothingness'' configuration
\bea
U_0(s) = u_0 \exp\le(-2 \al_2(s) P_0 - 3 \al_3(s)(P_0 \times P_0)\ri)~, \qquad \frac{d\al_i}{ds} = \lam_i(s)~. \label{notopen3}
\eea 
 Recall that any bulk spacetime of interest can be written in the form
\begin{align}
A & = L dL^{-1} \qquad L(x^{\pm}, \rho) = \exp(-\rho L_0) \exp\le(-\int_{x_0}^{x} dx^i a_i\ri)~, \\
\bA & = R^{-1} dR \qquad R(x^{\pm}, \rho) = \exp\le(\int_{x_0}^x dx^i \ba_i\ri) \exp\le(-\rho L_0\ri)~,
\end{align}
and hence our empty configuration simply becomes
\be
U(s) = L(s) U_0(s) R(s)~, \qquad P(s) = R^{-1}(s) P_0 R(s)~.
\ee

Demanding that the boundary condition \eqref{bcsl3} be obeyed we find the generalization of \eqref{rhseq} to the spin-$3$ case:
\be\label{eq:master}
\exp\le(-2 \Delta\al_2 P_0 - 3 \Delta \al_3(P_0 \times P_0) \ri) = (R(0) U_i^{-1} L(0))(R(s_f)U_f^{-1} L(s_f))^{-1}
\ee
All of the quantities on the right-hand side are known, and so we need simply solve this equation for $\Delta\al_2$ to determine the on-shell action. In practice solving this equation, even in a matrix representation,  can be somewhat cumbersome, and we will attempt to streamline the process as much as possible. The quickest route to the answer is to equate the eigenvalues of both sides. We have already found the eigenvalues of the left-hand side; they are simply $\exp\le(\lam_{\mathbb{P}}\ri)$, where the matrix of eigenvalues, $\lam_{\mathbb{P}}$, is given in \eqref{eigbigP} for the fundamental representation. 
 For notational convenience we denote 
\be
M \equiv (R(0) U_i^{-1} L(0))(R(s_f)U_f^{-1} L(s_f))^{-1}~, \label{Mdef}
\ee
and $\lam_M$ are the corresponding eigenvalues in the fundamental representation. We will set $U_i = U_f = 1$ for the reasons discussed in section \ref{sec:entgrav}.\footnote{We reiterate that  our results are sensitive to the choice of boundary conditions. It will be interesting to investigate the properties and physical interpretation of $W_{\cal R}(C)$ for $U_i \neq U_f \neq 1$.} As in \eqref{eq:alpha2},  we find the relation
\bea\label{eq:d2hs}
\Delta\al_2 = -\frac{1}{2\sqrt{2 c_2}}{\tr}_f(\log(\lam_{M}) L_0)
\eea

For the generic case the task of finding the eigenvalues of \eqref{Mdef} can be tedious (there are a handful of simple analytic cases which we will discuss below). Fortunately the process simplifies somewhat as we take the UV cutoff to infinity, i.e. the limit where $\ep^{-1} \equiv e^{\rho_0}$ is much greater than any other scale (e.g. the temperature, the inverse length of the interval, etc.). Now using \eqref{charpol3}, the characteristic polynomial of $M$ can be written 
\be
P_M(\lam_M) = -\lam^3_M + {\rm tr}_f(M) \lam^2_M - {1\over 2}({\rm tr}_f(M)^2 - {\rm tr}_f(M^2) ) \lam_M + 1 ~.
\ee
It is not difficult to solve this cubic equation. However, for the purpose of computing entanglement entropy we only need to know the behavior of the solutions for $\ep$ small.   Then via direct computation in all cases of interest we find that the traces of $M$ have a specific scaling with $\ep$ in the small $\ep$ limit.  It is easy to solve this equation in the small $\ep$ limit by picking pairs of terms and balancing their constituent terms against each other.
 Expanding in powers of $\ep$ we find:
\be
{\rm tr}_f(M) = \frac{m_1}{\ep^{4}} + \sO(\ep^{-2})~,\qquad {\rm tr}_f(M)^2 - {\rm tr}_f(M^2) = \frac{2 m_2}{\ep^{4}} + \sO(\ep^{-2}) ~.\label{m1m2def}
\ee
Here $m_1$ and $m_2$ are expansion coefficients that depend on the parameters of the problem. 

  We find the eigenvalues of $M$ to be
\be\label{eq:peeig}
\lam_{M} = \mbox{diag}\le(\frac{m_1}{\ep^4}, \frac{m_2}{m_1}, \frac{1}{m_2}\ep^4 \ri)~.
\ee
In the $\slt$ limit we have $m_1 = m_2$ as expected, and hence the eigenvalues are in the primary ordering. Now using \eqref{eq:d2hs} we can solve  for the on-shell action. The answer in the primary ordering is simply
\be
S_{\rm EE} = \sqrt{2 c_2}\log\le(\frac{\sqrt{m_1 m_2}}{\ep^4}\ri)~. \label{eem1m2}
\ee
 This is perhaps the most useful result of this section: we have reduced the problem of computing an entanglement entropy to computing traces of powers of the matrix $M$ defined in \eqref{Mdef}. We now present the results on various spacetimes of interest. 

\subsection{Gravitational sector of $\slh$}\label{sec:sl3grav}
As a warm up, and to illustrate some of the non-trivial structure in the equations, let's consider cases solving \eqref{eq:master} for the gravitational subsector.  That is, consider connections $(A,\bar A)$ for which 
\be\label{eq:subsub}
L(s)~,~R(s)\in \slt\subset \slh ~,
\ee
where the  $\slt$ subgroup  can be either the one characterizing the principal or diagonal embeding.  Again we will use $U_i = U_f = {\bf 1}$ as our boundary conditions. Then the right hand side of \eqref{eq:master} belongs to the  $\slt$ subgroup as well. This imposes a non-trivial constraint on $P_0$ and in particular it implies that
\be\label{eq:posim}
-2 \Delta\al_2 P_0 - 3 \Delta \al_3(P_0 \times P_0) \in sl(2,\RR)~.
\ee
Given the constraints on the quantum numbers of our probe---i.e. having $h\neq0$ and $w=0$---it is natural to choose $P_0\in  sl(2,\RR)$. This will further force $\Delta\al_3=0$ in  \eqref{eq:posim}. With the simplification $\Delta\al_3=0$, it is then clear that  for any connection in the class \eqref{eq:subsub} the analysis reduces again to our discussion in section \ref{sec:openI}. 

However, even in the subclass \eqref{eq:subsub}, there are as well solution to \eqref{eq:master} which have $P_0\notin  sl(2,\RR)$ while still satisfying  $c_2\neq0$, $c_3=0$ and \eqref{eq:posim}.  A simple computation will show that these other configurations correspond to probes which have both $h$ and $w$ satisfying $h=\pm w/2$ and $c_3=0$ in accordance to \eqref{cascharges}. While these probes are still physical, and rather interesting, they change the representation ${\cal R}$; and for the purposes of making a comparison with known results in the dual CFT,   it  is not the appropriate choice of quantum numbers.  This is exactly the same phenomena we encountered when computing thermal entropy. The different ordering of the eigenvalues, such as the one illustrated  in  \eqref{eigbigPnew}, gave distinct solutions to the equations of motion. This  changes the value of the on-shell action, but most importantly we emphasize that it modifies the representation of the algebra which defines the Wilson line. 

As a final remark, for these simple  backgrounds \eqref{eq:subsub},  the primary ordering in \eqref{eq:peeig} is the only solution which is compatible with the condition $P_0\in sl(2,\RR)$, and hence having a probe with vanishing $w$ charge. As it will be clear in the following examples, \eqref{eem1m2} is in complete agreement with the results in section \ref{sec:openI}.

\subsection{Diagonal embedding}
We begin with the diagonal embedding. As a warmup we compute the entanglement entropy of an open interval in the diagonal embedding AdS$_3$ vacuum, which is given by the very simple connections
\be
a = W_2 dx^+~, \qquad \ba = W_{-2} dx^-~.
\ee
In this case $M$ is easily explicitly evaluated and we find $m_1$ and $m_2$ to be
\be
m_1 = (\Delta\phi)^2 ~,\qquad m_2 = (\Delta\phi)^2 \ . \label{diagAdS3}
\ee
Evaluating \eqref{eem1m2} and using \eqref{casPD} to fix $c_2$ we find an entanglement entropy of
\be
S_{\rm EE} = \frac{c_D}{3} \log\le(\frac{\Delta\phi}{\ep^2}\ri),
\ee
where $c$ is the central charge of the CFT in the diagonal embedding. This is of course the expected result from 2D CFT, but it is somewhat gratifying to see it emerge from a higher-spin computation. 

It is only slightly more difficult to compute the entanglement entropy for the black hole in the diagonal embedding; in this case the relevant connections were given in \eqref{diagBH}. Evaluating the relevant traces we find
\begin{align}
m_1 & = \frac{1}{4\om}\exp\le(-\frac{2}{3}(2q + \eta + 12 \sqrt{\om})\Delta\phi\ri)(e^{8 \sqrt{\om}\Delta \phi} - 1)^2~,  \\
m_2 & = \frac{(e^{8 \sqrt{\om}\Delta\phi} - 1)^2 \exp\le(\frac{2}{3}(2q + \eta - 12 \sqrt{\om})\Delta\phi\ri)}{4\om}~,
\end{align}
leading to the entanglement entropy
\be\label{eq:opendiag}
S_{\rm EE} = \frac{c_D}{3} \log\le(\frac{1}{\sqrt{\om} \ep^2} \sinh\le(4 \sqrt{\om} \Delta \phi\ri)\ri) \ .
\ee
This is the familiar expression for an entanglement entropy in 2d CFT at finite temperature, and it only depends on the spectral flow invariant $\om$ (see \cite{Ammon:2011nk,Castro:2011fm}). This came from a covariant higher-spin calculation; however the final answer is equal to that arising from the usual Ryu-Takayanagi prescription in the metric representation of the theory (this is analagous to the fact that the thermal entropy \eqref{diagtherm} is equal to that predicted by the usual Bekenstein-Hawking formula). The presence of the nontrivial $U(1)$ holonomies does not change this result. 
 
To end this subsection, we would like to solve \eqref{eq:master} for the background \eqref{diagBH} using the same logic as in the previous section \ref{sec:sl3grav}. Notice that $W_0$ commutes with $\{L_0,W_{\pm 2}\}$, so for the diagonal embedding black hole \eqref{diagBH} the matrix $M$ can be decomposed as
\be
M= M_{SL(2)} M_{W_0}~,
\ee
where $M_{SL(2)}$ contains the exponentials of $\{L_0,W_{\pm 2}\}$  and $M_{W_0}$ the contribution to $M$ from $(a,\bar a)$ which depend on $W_0$. Next if we set $P_0\in sl(2,\RR)$, then equation \eqref{eq:master} reduces to
\be
\exp\le(-2 \Delta\al_2 P_0 \ri) =  M_{SL(2)} ~,\quad \exp\le(- 3 \Delta \al_3(P_0 \times P_0) \ri) = M_{W_0}~.
\ee
  Using the fundamental representation of the $\slh$ matrices it is not difficult to solve for $\Delta \alpha_i$. It also makes clear that $\Delta\al_2$ will only depend on $\om$ and be insensitive to $(q,\eta)$ as it is in \eqref{eq:opendiag}. But if 
  $P_0\notin sl(2,\RR)$ we end up with a probe with non-vanishing $W_0$ charge, which corresponds to an different ordering of the eigeinvalues in \eqref{eq:master}.
  

\subsection{Principal embedding}\label{sec:openPEBH}
We turn now to the principal embedding. We begin with an open interval in the principal embedding AdS$_3$ vacuum, given by the simple connections:
\be
a = L_1 dx^+~, \qquad \ba = -L_{-1} dx^-~,
\ee
from which we find
\be
m_1 = (\Delta\phi)^4 ~,\qquad m_2 = (\Delta\phi)^4 \ . \label{princAdS3}
\ee
Again evaluating \eqref{eem1m2} and using \eqref{casPD} to relate $c_2$ to the principal embedding central charge we find the entanglement entropy to be
\be
S_{\rm EE} = \frac{c_P}{3} \log\le(\frac{\Delta \phi}{\ep}\ri) \ .
\ee
We again find the expected result from 2d CFT. Note that though the scaling of $m_1$ and $m_2$ with $\Delta\phi$ is different in the two embeddings, the corresponding relations between $c_2$ and the central charges also differ in just the right way to give the correct prefactor. 

Next, we study the black hole in the principal embedding, where the connection was given previously in \eqref{princBH}. The evaluation of  $M$ in this case is somewhat more difficult due to the increased complexity of $a$, $\ba$. The resulting exact expressions for $m_1$ and $m_2$ are somewhat lengthy sums of exponentials. We discuss their derivation in Appendix \ref{app:princ}. In this section we discuss only the asymptotic expressions. 

For very small interval length compared to the temperature we find
\be
m_1\le(\Delta\phi \ll \beta\ri) = m_2\le(\Delta\phi \ll \beta\ri) \sim \frac{9 C }{4(3-2 C)^2}\le(\sqrt{\frac{k}{2\pi \sL}}\ri)^4 \Delta\phi^2 \ . \label{UVscalP}
\ee
Similarly, for very large interval length we find
\begin{align}  
m_1\le(\Delta\phi \gg \beta\ri) & \sim  b_1 \exp\le(4\sqrt{\frac{2\pi \sL}{k}}\Delta\phi\le(\frac{3 + C(\sqrt{4 C - 3}-2)}{\sqrt{C}(2 C - 3)}\ri)\ri)~, \\   
m_2\le(\Delta\phi \gg \beta\ri) & \sim  b_2 \exp\le(4\sqrt{\frac{2\pi \sL}{k}}\Delta\phi \le(\frac{-3 + C( \sqrt{4 C - 3}+2)}{\sqrt{C}(2C - 3)}\ri)\ri)~,
\end{align}
where $b_1$ and $b_2$ are calculable but uninteresting functions of $C$. We first focus on the infrared limit: evaluating \eqref{eem1m2} at large $\Delta\phi$ we find the entanglement entropy to be:
\be
S_{\rm EE}\le(\Delta\phi \gg \beta\ri) \sim \Delta\phi\le(2 \sqrt{2 \pi \sL k}\frac{\sqrt{1 - \frac{3}{4C}}}{1 - \frac{3}{2 C}}\ri)~.
\ee
Here we have used \eqref{casPD} to set $\sqrt{2 c_2} = \frac{c_P}{12} = \frac{k}{2}$. The entropy density extracted from this asymptotic expression precisely agrees with the thermal entropy calculated previously in \eqref{thermprinc} (recall in that expression we assumed that $\phi$ has a periodicity of $2\pi$, whereas here we are allowing it to be noncompact). This agreement between entropy densities is clearly required for consistency, but it is not an obvious identity in our formalism.  

We turn now to the small $\Delta\phi$ limit given by \eqref{UVscalP}. It is important to note that for finite $C$ the small $\Delta\phi$ limit of these expressions no longer agrees with the principal embedding AdS$_3$ vacuum \eqref{princAdS3}; it appears that the UV structure of the theory is different. This is expected: to obtain a black hole that carries the higher spin charge, we have applied a higher spin chemical potential, which is a deformation of the CFT Lagrangian by a dimension $3$---and thus irrelevant---operator \cite{Gutperle:2011kf, Ammon:2011nk}. {This operator modifies the theory in the UV: in fact in the UV the theory flows to the diagonal embedding vacuum, which is consistent with the fact that the small $\Delta\phi$ scaling appearing in \eqref{UVscalP} is that of the diagonal embedding AdS$_3$ vacuum \eqref{diagAdS3}}.\footnote{ As shown in \cite{Compere:2013gja}, for finite (non-zero)  and constant values of $\mu$ there exists a consistent set of boundary conditions  which preserve the ${\cal W}_3$ symmetry; a modification of the boundary conditions will give ${\cal W}_3^2$ symmetry. This makes it rather unclear what the appropriate or natural boundary conditions are to describe the UV theory, and it might be relevant for the discussion in section \ref{sec:RGDI}. } 

In the next subsection we address some features of this flow.

\subsection{RG flow from diagonal to principal}\label{sec:RGDI}
There is a simpler bulk connection which captures the physics of the flow from the diagonal embedding to the principal embedding \cite{Ammon:2011nk}:
\be
a = \hat\lam L_1 dx^+ + \frac{1}{4} W_2 dx^- ~,\qquad \ba = -\hat\lam L_{-1} dx^- + \frac{1}{4} W_{-2} dx^+ \ ,
\ee
with $\hat\lam$ a constant (to not be confused with the Lagrange multipliers or an eigenvalue). It is instructive to examine the full bulk connections $A, \bA$:
\be
A = \hat\lam e^{\rho}L_1 dx^+ + e^{2\rho} \frac{W_2}{4}dx^-  + L_0 d\rho~, \qquad \bA = -\hat\lam e^{\rho} L_{-1} dx^- + e^{2\rho} \frac{W_{-2}}{4} dx^+ - L_0 d\rho~.
\ee
We see that at large $\rho$ (i.e. the UV) the connection is that of the diagonal embedding vacuum, where $\slt$ is generated by $W_{\pm 2}, L_0$. At small $\rho$ (i.e. the IR) it crosses over to that of the principal embedding vacuum, where $\slt$ is generated by $L_{\pm 1}, L_0$. The parameter $\hat\lam$ governs the size of this domain wall solution.  The principal embedding black hole may be viewed as a finite-temperature generalization of this RG flow. From the point of view of the IR principal embedding theory there is a spin-$3$ deformation turned on that takes the theory to the diagonal embedding, as described above.   

It is thus interesting to compute the entanglement entropy on this background. We find
\be
m_1 = (\Delta\phi)^2(1 - \hat \lam^2\Delta\phi)^2~, \qquad m_2 = (\Delta\phi)^2(1 + \hat \lam^2\Delta\phi)^2~.
\ee
Note that these values interpolate between the diagonal embedding results at small $\Delta\phi$ and the principal embedding results at large $\Delta\phi$. Evaluating this directly we find for the entanglement entropy:
\be
S_{\rm EE} = \frac{c_D}{6} \log\le(\frac{\Delta\phi^2|1- \hat \lam^4 \Delta\phi^2|}{\ep^2}\ri), \label{RGflowans}
\ee
where we have used the value of $\sqrt{c_2}$ appropriate to the diagonal embedding vacuum \eqref{casPD}. There are some curious features in this expression. First, while we recover the correct entanglement entropy in the UV, for $\Delta\phi \gg \hat\lam^{-2}$ we find
\be
S_{\rm EE}(\Delta\phi \to \infty) \sim \frac{2 c_D}{3} \log \Delta\phi
\ee
suggesting an effective IR central charge $c_{IR} = 2 c_D$. This is not what we would naively expect, as in reality the central charge of the principal embedding vacuum (which is expected to govern the infrared physics) is related to that of the diagonal embedding by $c_P = 4 c_D$. This discrepancy occurred because we picked the value of $c_2$ to be appropriate to the diagonal embedding vacuum: if we had picked it to be appropriate to the principal embedding vacuum we would have gotten the right answer in the IR but not in the UV. We find it somewhat perplexing that our formalism does not  allow us to cross through this RG flow. 

\begin{figure}[h]
\begin{center}
\includegraphics[scale=0.6]{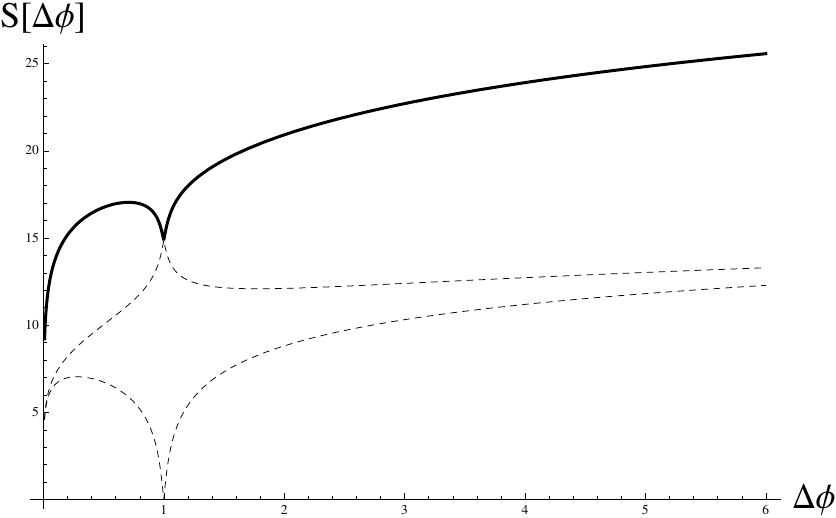}
\end{center}
\vskip -0.5cm
\caption{Solid line is entanglement entropy in primary ordering with $\hat\lam = 1$. In the vicinity of the singular region we have used the exact eigenvalues of $M$; we are not using the scaling assumed in $\eqref{m1m2def}$. Note sharp singularity at $\Delta\phi = 1$. Dashed lines are non-primary orderings, shown for illustrative purposes. }
\label{fig:openintRG}
\end{figure}

Next we note a more perplexing fact still: the argument of the logarithm vanishes at $\Delta\phi = \pm \hat\lam^{-2}$. Naively speaking this appears to imply the nonsensical result that the entanglement entropy is arbitrarily negative: in reality what is happening is that either $m_1$ or $m_2$ vanishes, and thus the dependence on the UV cutoff assumed in \eqref{m1m2def} is breaking down. If we perform the whole calculation without assuming that scaling, then this divergence is regularized by the cutoff, as shown in Figure \ref{fig:openintRG}; nevertheless  we find it peculiar that the UV cutoff manifests itself in an unexpected way here at intermediate scales. Furthermore, at this point we find that a non-primary ordering is becoming degenerate with the primary ordering.\footnote{Note that on either side of the singularity an ordering is fixed by demanding continuity with an $\slt$ limit. Furthermore a small variation of the boundary conditions on the probe in the direction $U_i \sim e^{\al W_0}$ lifts 
this 
degeneracy, separating the two curves (it also smoothens the singularity at $\Delta\phi = \pm \hat\lam^{-2}$). Thus despite the temptation to move from the solid to the dashed line we do not believe this is physically appropriate, and that the solid line is the correct answer for all $\Delta\phi$.} This appears to be an interesting manifestation of higher spin physics, but we must admit that at the moment we are not certain what this indicates physically. 

\begin{figure}[h]
\begin{center}
\includegraphics[scale=0.6]{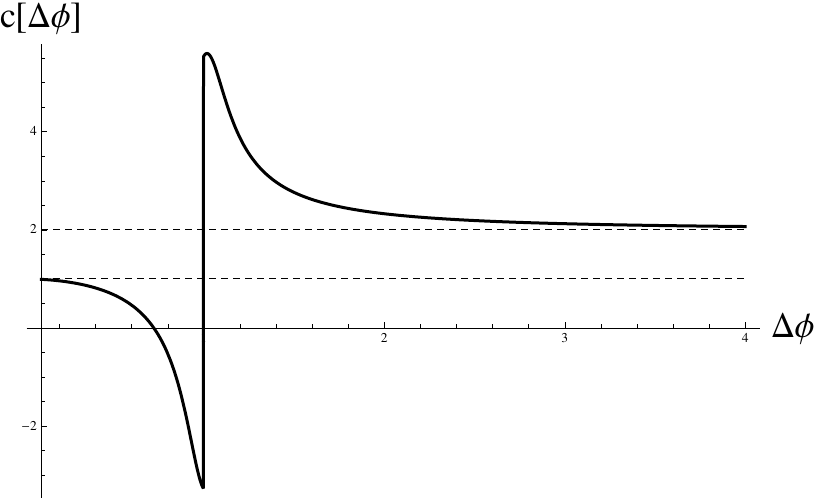}
\end{center}
\vskip -0.5cm
\caption{Plot of entropic $c$-function, normalized so $c(\Delta\phi \to 0)/c_{D}=1$. Dashed lines indicate UV and IR asymptotic values; note effective central charge increases only by a factor of $2$. The negative values attained by the c-function and its singularly positive jump violate strong subadditivity. The height of the jump is non-universal, depending on the UV cutoff.}
\label{fig:cfunc}
\end{figure}

To further attempt to interpret the singularity it is convenient to introduce the entropic $c$-function \cite{Casini:2004bw}:
\be
c(\Delta\phi) \equiv \Delta\phi S'(\Delta\phi) \ .
\ee
This can be thought of as a measure of the number of degrees of freedom at the length scale corresponding to $\Delta\phi$: at a conformal fixed point it directly measures the central charge. We plot this quantity for the primary ordering in Figure \ref{fig:cfunc}. The singularity at $\Delta\phi = \hat{\lam}^{-2}$ is a discontinuous jump of $c(\Delta\phi)$ and so may be interpreted as the singular injection of new degrees of freedom at this scale. The height of this jump is non-universal: it diverges as the UV cutoff is taken to infinity. Downwards jumps in the $c$-function (or its appropriate higher-dimensional generalization \cite{Liu:2012eea}) have been noted before in the context of first-order phase transitions in the holographic entanglement entropy along RG flows \cite{Myers:2012ed,Liu:2012eea,Klebanov:2007ws,Pakman:2008ui}. Our jump differs from these previous examples in that its magnitude depends on the UV cutoff and further in that it is a jump {\it upwards}.

Finally, we turn to the issue of strong sub-additivity of the entanglement entropy. This is a powerful set of constraints on entanglement entropy in general quantum mechanical systems that require little input besides basic data about Hilbert spaces. In static situations, the Ryu-Takayanagi prescription automatically satisfies strong sub-additivity \cite{Headrick:2007km}, but as we will see, our formalism need not. In particular, strong sub-additivity of the entanglement entropy imposes constraints on the possible dependence of $S_{\rm EE}$ on $\Delta\phi$; as shown in \cite{Callan:2012ip}, for a single interval in an infinite system with one spatial dimension, these constraints are
\be
S_{\rm EE}'(\Delta\phi) > 0~, \qquad S''_{EE}(\Delta\phi) < 0~.  \label{ssacons}
\ee
For small $\Delta\phi$, we see from Figure \ref{fig:openintRG} that the entanglement entropy is not monotonically increasing and the first of these two expressions is not satisfied. The positive discontinuity in the $c$-function can be viewed as a delta-function violation of the second expression.\footnote{It is interesting to note that the two conditions \eqref{ssacons} do not themselves rule out a sufficiently {\it gentle} increase in the entropic $c$-function: to exclude this one requires the further input of Lorentz invariance \cite{Casini:2004bw}. However the discontinuous upwards jump of our $c$-function clearly does violate \eqref{ssacons}.} Of course, for small $\Delta\phi$, the diagonal embedding vacuum is also not unitary \cite{Castro:2012bc}: thus the presence of negative-norm states means that we cannot be sure the reduced density matrix has no negative eigenvalues, and so we violate one of the only conditions in the proof for strong sub-additivity. We are somewhat reassured that at large $\Delta\phi$ (where we have presumably flowed to the unitary principal embedding), both of these constraints {\it are} satisfied. 

{Thus from the point of view of our proposal of entanglement entropy we see that this RG flow background is somewhat puzzling.} We believe there is more to be understood here, and at the moment we are not certain whether our proposal should be modified in some way or if these features are instead a manifestation of peculiarities associated with the RG flow itself. We discuss two such pathologies below:
\ben 
\item As we have emphasized, the UV theory---the diagonal embedding---is not unitary and thus we are not certain that our intuition regarding RG flows should apply here. For example note that the central charge {\it increases} along the flow; while this is not expressly forbidden by any $c$-theorem (as the flow is not Lorentz-invariant and the UV theory is non-unitary anyway) we still find it somewhat distressing. 
\item The flow is also somewhat peculiar kinematically; to understand this consider the correlator of the diagonal embedding stress tensor along the flow, which was worked out in \cite{Ammon:2011nk}:
\be
\langle T(p)T(-p) \rangle = \frac{p_+^3 p_-}{\hat\lam^4 - \frac{4}{3} \frac{p_+^4}{p_-^2} }~.
\ee
The correlator has two distinct kinematic regimes; if $\frac{p_+^2}{p_-^2} \gg \hat\lam^4$ then $\langle T T \rangle \sim \frac{p_-^3}{p_+}$, corresponding to an operator of dimension $(2,0)$. This is meant to be the UV, as it corresponds to high $p_+$, and indeed this is the correct dimension for the stress tensor. In the other kinematic regime we have $\frac{p_+^2}{p_-^2} \ll \hat\lam^4$, leading to a correlator scaling as $\frac{p_+^7}{p_-}$ corresponding to an operator of dimension $(0,4)$. This is usually understood as the IR, as it appears to correspond to low $p_+$. However, it is possible to take $p_+ \to 0$ and still remain in the ``UV'' region, provided we also take $p_-$ to $0$ as well in an appropriate way. One can be at low momentum and still probe UV physics; inverting the logic, one can also be at high momentum and still probe IR physics. Thus the RG flow is kinematically rather different from most RG flows that we are familiar with. This is related to the Lorentz-breaking character of the 
deformation.\footnote{We briefly digress and discuss an example where a similar phenomenon happens: consider a Fermi surface, which may also be obtained from a Lorentz-invariant theory of relativistic fermions by a Lorentz-breaking deformation $\mu \bpsi \gamma^0 \psi$. There as well we find a similar breakdown: we can be at high momentum $k \to k_F \neq 0$ yet still probe IR physics due to the existence of gapless modes at the Fermi surface. It is also well-known that the Fermi surface state has peculiar entanglement entropy properties, including a logarithmic violation of the area law \cite{Wolf:2006zzb}. The pathologies in our system are of course quite different, but it is interesting to speculate whether they are also related to a breakdown of familiar kinematics.}
\een
These peculiarities are actually hidden in our calculation as well: for example, the result for the entanglement entropy \eqref{RGflowans} is quite sensitive to the reference frame. If we instead consider an open interval in a state that is slightly boosted compared to the rest frame induced by the deformation, we find a rather different and quite complicated answer which deserves to be better understood. 
\section{Conclusion} \label{sec:conc}

This has been a long journey: here we summarize the key points from our analysis. We have proposed that in theories of higher spin gravity there is a particular choice of open-ended bulk Wilson line that computes entanglement entropy in the dual field theory: 
\be
S_{\rm EE} = -\log\le(W_{\sR}(C)\ri)~,
\ee 
where the representation $\sR$ is a particular infinite-dimensional highest weight representation of the bulk gauge group $SL(N,\mathbb{R})$ and the trajectory $C$ extends into the bulk. We developed a considerable amount of technology to evaluate these Wilson lines, representing them as path integrals over an auxiliary worldline field called $U(s)$, whose action we showed to be (in the $\slh$ case):
\be
S(U,P; A,\bA)_C = \int ds \le( \Tr(P U^{-1}D_{s} U) + \lam_2 (\Tr(P^2) - c_2) + \lam_3 (\Tr(P^3) - c_3)\ri)~. \label{probefinal}
\ee
We then evaluated these path integrals in a saddle point approximation, developing techniques for solving the equations of motion of $U(s)$ and finding the on-shell action in terms of data specifying the bulk gauge connections. 

As the theory is topological, the actual trajectory that $C$ takes in the bulk does not matter; the answer depends only on data specified at the endpoints and can be expressed in a fairly explicit form, e.g. as in \eqref{eem1m2}. In the case of $\slt$ we demonstrated explicitly that this bulk Wilson line computes a conventional proper distance: indeed, we even showed that the usual geodesic equation made a somewhat unexpected appearance. In the case of $\slh$ gravity this Wilson line provides a gauge-invariant generalization of the formula of Ryu and Takayanagi, and we argued that for an appropriate choice of quantum numbers of the probe (i.e. the right choice of $c_2$ and $c_3$ above) it implements the replica trick approach to computing entanglement entropy in the dual field theory. 

We further showed that the computation of this quantity for a single interval often (but not always) results in sensible answers, i.e.
\ben
\item We reproduced expected results for entanglement entropy that are fixed by conformal invariance; in our approach however these all involve only an $\slt$ subgroup of the full higher spin gauge group.
\item When the Wilson line is looped around a black hole horizon, our proposal computes thermal entropies which depend nontrivially on higher spin charges: we demonstrated equivalence with a general formula for thermal entropies computed from a Euclidean action by \cite{deBoer:2013gz}. 
\item In some cases (e.g. the diagonal embedding black hole with nontrivial holonomies) our proposal recovers expected but nontrivial dependence of the entanglement entropy on the interval length in a manifestly gauge-invariant manner. 
\een
However, we are somewhat reluctant to claim complete victory, as we also found some puzzling features:
\ben

\item There are currently two expressions in the literature for the entropy of the principal embedding black hole, which are thought to differ due to modified expressions for the energy of the dual field theory and thus different entropies as computed from the first law of thermodynamics. Our analysis agrees with one of them, but we do not have a proposal which will reproduce the other entropy. This other thermal entropy is in agreement with CFT calculations \cite{Kraus:2011ds,Gaberdiel:2012yb}, hence it is important to relate it to entanglement entropy computations.

\item Our analysis of a particular non-unitary RG flow geometry failed to demonstrate the expected variation in the central charge from UV to IR, violates strong sub-additivity, and displayed a peculiar singularity at intermediate distance scales. At the moment we are unable to explain these features, and are unsure as to whether our proposal must be modified or whether we are accurately computing an entanglement entropy that simply behaves oddly due to unwanted pathologies of the RG flow geometry, such as its lack of unitarity. Further computations on related backgrounds should be helpful in elucidating this. 
\een

Thus we believe there is still a great deal to be understood here. In addition to the resolution of the above two puzzles, some other concrete directions for future research are:
\ben
 
\item We have not systematically explored the space of boundary conditions on our probe. In particular, the choice that we have made, that $U(0) = U(s_f) = {\bf 1}$, was motivated by analogy with the $\slt$ case rather than from a systematic study of its true meaning in the higher spin case. It is quite possible that there are other sensible choices which may compute a different class of observables. It would be very useful to understand a principle that would help us choose between them. A proper quantum treatment of the action \eqref{probefinal} (which does exist for the $\slt$ case \cite{Dzhordzhadze:1994np}) may help with this, as will a more refined understanding of the backreaction of the probe on the bulk geometries in question. 

\item Our treatment fixes the Casimirs of the representation; as we have explained, these Casimir constraints can be solved by a discrete number of inequivalent representations which are visible as unwanted solutions to our equations of motion. It is desirable to find a more sophisticated treatment of the representation that would not contain these spurious solutions. 

\item In all of our calculations we have frozen the bulk gauge connections and viewed this Wilson line as a probe. A rough justification for this was given back in \eqref{c2nrel}, where we showed that in the limit that we are computing an entanglement entropy, the effective $c_2$ of the probe vanishes as we take $n \to 1$. If we were interested in computing instead a Renyi entropy, we would have a finite $n$ and thus backreaction on the geometry would be significant: in fact at integer $n$ the bulk geometries are regular and very different from empty space with a conical singularity. These bulk geometries have been constructed in the metric representation of the $\slt$ case in \cite{Faulkner:2013yia}; it would be very interesting to understand a Chern-Simons representation of those geometries, perhaps with the help of our Wilson lines, opening the door to the construction of Renyi entropies in higher spin theories. 

\item Our discussion presents a new way to think about the Ryu-Takayanagi formula and may admit further generalizations to other theories of 3d gravity with a Chern-Simons description, e.g. to topologically massive gravity. In this paper we have focused on $\slh$; however if we are systematic we cannot see any real obstacles to extending the formalism to any finite $N$ in $SL(N, \mathbb{R})$, and with more work perhaps even to the infinite dimensional Lie algebra $hs[\lambda]$. In the latter case the precise understanding of the dual field theory might allow an interpolation between the gravitational and field-theoretical ways of thinking about entanglement entropy. In many ways, this would provide a concrete realization of the idea of the {\it construction} of bulk geometry from entanglement that we alluded to in the introduction to this paper. 
\een

We hope to return to some of these issues in the future.

To conclude, our proposal provides a reformulation of the ideas of holographic entanglement entropy, one that takes seriously the topological nature of 3d gravity and permits both concrete calculations and further generalizations. We hope that it may play a small role in understanding the interplay of classical geometry and quantum entanglement. 

\begin{acknowledgements} 

It is a pleasure to acknowledge helpful discussions with T. Azeyanagi, J. de Boer, A. Campoleoni, G. Compere, C. Cordova, S. Detournay, M. Gutperle, M. Headrick, T. Faulkner, P. Kraus, K. Jensen, J. I. Jottar, V. Kumar, R. Loganayagam, D. Morrison, C. Peng, E. Perlmutter, W. Song and D. Stanford. We are grateful to the KITP program on Bits, Branes, and Black Holes at which this work was initiated back in the mists of antiquity. We thank the IoP at the University of Amsterdam and  Galileo Galilei Institute for Theoretical Physics for their hospitality, and the INFN for partial support during the completion of this work. MA was supported in part by the National Science Foundation under Grant No. NSF PHY-07-57702. AC's work is  supported by the Fundamental Laws Initiative of the Center for the Fundamental Laws of Nature, Harvard University. NI is supported in part by the National Science Foundation under Grant No. PHY11-25915.

\end{acknowledgements}

\begin{appendix}

\section{Conventions and useful relations}\label{app:conv}

\subsection{Elementary matrix algebra}\label{app:matrix}

Here we summarize some basic matrix identities that we use in explicit computations above. For any matrix $A$, the characteristic polynomial $P_A(\lam_A)$ of $A$ can be written in terms of traces of powers of $A$. Denote $A_n \equiv \Tr(A^n)$ and write
\be
P_A(\lam_A) = \det(A - \lam_A {\bf 1}) = \exp\le[\Tr \log( A - \lam_A {\bf 1})\ri] \label{detform}
\ee
Now Taylor expanding the right hand side in powers of $\lam_A$, we can find an explicit expression for the characteristic polynomial; e.g. if $A$ is $3 \times 3$ we have
\bea
P_A(\lam)& =& -\lam^3_A + A_1 \lam^2_A - \ha \lam_A(A_1^2 - A_2) - \frac{1}{6}\le(-A_1^3 + 3 A_1 A_2 - 2 A_3\ri)\cr
& =& -\lam^3_A + A_1 \lam^2_A - \ha \lam_A(A_1^2 - A_2) +\det(A)~. \label{charpol3}
\eea
Note that there are infinitely many terms in the expansion of the right-hand side of \eqref{detform} in powers of $\lam_A$; as we know that the left-hand side is a simple polynomial in $\lam_A$, all but the first three of them must be identically $0$. The relations that enforce this allow one to determine traces of arbitrarily high powers of $A$ in terms of traces of the first three powers.

\subsection{$SO(2,2)$ and $SL(2,\RR)$ conventions}

Our conventions for the $so(2,2)$ algebra are 
\be [M_a,M_b]=\epsilon_{abc}M^c~,\quad [M_a,P_b]=\epsilon_{abc}P^c~, \quad [P_a,P_b]=\ell^2 \epsilon_{abc}M^c~.\ee
Alternatively, if we define $J_a^\pm={1\over 2}(M_a\pm \ell P_a)$ the commutators simplify to
 \be
 [J_a^+,J_b^+]=\epsilon_{abc}J^{+c}~,\quad [J_a^-,J_b^-]=\epsilon_{abc}J^{-c}~,\quad [J_a^+,J_b^-]=0~,\ee
where we are decomposing $so(2,2)=sl(2,\RR)_L\times sl(2,\RR)_R$ and  $J^\pm_a\in sl(2,\RR)_{L,R}$.

In general, we denote the 3 generators of $sl(2,\RR)$ as $\{J_0,J_{1},J_{-1}\}$. And in accordance with the above, the algebra is given by
\be
[J_a,J_b]= \epsilon_{abc}J^c ~,\quad J^c=\delta^{cd}J_d~,
\ee
where $\epsilon_{abc}$ is a completely antisymmetric tensor and $\epsilon_{0+-}=1$. The metric is given by 
\be\label{eq:delta}
\delta_{00}={1\over 2}~,\quad \delta_{+-}=\delta_{-+}=-1~.
\ee
The inner product  is defined as
\be
p^ap_a=p^ap^b\delta_{ab}={1\over 2}p^0p^0-2p^+p^-~.
\ee

\subsubsection{Fundamental representation}
\be
J_0= \left[\begin{array}{cc}1/2 & 0 \\ 0& -1/2\end{array}\right]~,\quad J_{1}= \left[\begin{array}{cc}0& 0 \\ -1& 0\end{array}\right]~,\quad J_{-1}= \left[\begin{array}{cc}0 & 1 \\ 0& 0\end{array}\right]~.
\ee
For this representation, it is straight forward to show that
\be
J_aJ_b={1\over 2}\delta_{ab}+{1\over 2}\epsilon_{abc}J^c~,
\ee
and $\tr_f(J_aJ_b)=\delta_{ab}$. From here we can derive some useful properties. For example,
\be
\exp(\kappa p^aJ_a)=\cosh(\kappa) 1_{2\times 2}+ \sinh (\kappa)  p^aJ_a ~, \quad  {\rm if}~~ p^ap_a=2~,
\ee
and
\be
\exp(\kappa p^aJ_a)=1_{2\times 2} + \kappa  p^aJ_a ~, \quad {\rm if} ~~ p^ap_a=0~.
\ee

\subsubsection{Adjoint representation}
\be
J_0= \left[\begin{array}{ccc}0&0&0\\ 0&1 & 0 \\ 0& 0& -1\end{array}\right]~,\quad J_{1}= \left[\begin{array}{ccc}0& 1&0 \\ 0& 0&0\\ 2&0&0\end{array}\right]~,\quad J_{-1}= \left[\begin{array}{ccc}0& 0& -1 \\ -2&0& 0\\ 0&0&0 \end{array}\right]~,
\ee
and
\be
 {\rm tr}_{\rm adj}(J_a J_b)=4 \delta_{ab}~,
\ee
with $\delta_{ab}$ given by \eqref{eq:delta}.
In this representation we have $J_0^3=J_0$ and $J_{\pm1}^3=0$ and hence we find
\bea\label{eq:expa}
e^{\kappa J_{\pm}}&=&1_{3\times 3} + \kappa  J_{\pm 1}+{\kappa^2\over 2}   J_{\pm 1}^2~,\cr
e^{\kappa J_0}&=&1_{3\times 3} +(\cosh( \kappa)-1) J_0^2 +\sinh(\kappa) J_0
\eea
In general a $3\times 3$ traceless matrix $X$ satisfies
\be\label{eq:x33}
X^3 = {1\over 3} {\rm tr}(X^3) 1_{3\times 3} +{1\over 2}\tr (X^2) X
\ee
And if $X\in SL(2,\RR)$ we further have that ${\rm tr}(X^3)=0$. Written differently 
\bea
(p^a J_a )^3 &=& {1\over 2} p^a p^b \Tr(J_a J_b) p^c J_c  \cr
&=& 2 p^a p_a p^c J_c 
\eea
Final property is the generalization of \eqref{eq:expa}
\be
\exp(\kappa p^a J_a)=1_{3\times 3}+(\cosh(\kappa)-1)(p^aJ_a)^2 + \sinh(\kappa) p^aJ_a~,
\ee
for  $2p^ap_a=1$.

\subsection{$SL(3,\RR)$ conventions} \label{app:sl3}

We label the $sl(3,\RR)$  generators as $T_a=\{L_i,W_m\}$ with $i=-1,0,1$ and $m=-2,\ldots,2$. The algebra reads
\bea
[L_i ,L_j] &=& (i-j)L_{i+j}~, \cr
[L_i, W_m] &=& (2i-m)W_{i+m}~, \cr
[W_m,W_n] &=& -{1 \over 3}(m-n)(2m^2+2n^2-mn-8)L_{m+n}~.
\eea

The algebra has two independent Casimirs which are defined as
\be\label{eq:c2c3}
C_2 = \delta^{ab} T_a T_b ~,\quad C_3 = h^{abc}T_aT_bT_c~.
\ee 
Here $\delta_{ab}$ and $h_{abc}$ are symmetric tensors which define the Killing forms of the Lie Algebra. Indices are raised using $\delta^{ab}$ where $\delta^{cb}\delta_{ab}={\bf 1}_{8\times 8}$. These tensor are given by
\be
\delta_{ab}= {\rm tr}_f (T_a T_b) ~,\quad h_{abc}={\rm tr}_f(T_{(a}T_bT_{c)})~.
\ee
The overall normalizations of this tensor is ambiguous, so in the definition above we fixed this ambiguity by using the matrix trace in the fundamental representation. We could have used instead e.g. the adjoint representation,  which would modify certain normalizations and definitions in the text---but of course the physics is  unchanged given that conventions are implemented consistently.

In the text we use the short hand notation $\Tr(P^n)$ which should be read as contractions with the Killing forms, i.e.
\be\label{eq:deftr}
\Tr(P^2)= P^aP^b\delta_{ab}~,\quad \Tr(P^3)= P^aP^bP^c h_{abc}~.
\ee

\subsubsection{Fundamental representation}
We work with the following matrices in the fundamental representation
\bea
L_1 & =& \bmat 0&0&0 \\ 1&0&0 \\ 0&1&0\emat,\quad L_0= \bmat 1&0&0 \\ 0&0&0 \\ 0&0&-1\emat,\quad  L_{-1} = \bmat 0&-2&0 \\ 0&0&-2 \\ 0&0&0\emat~,\cr &&\cr
W_2 &= &2 \bmat 0&0&0 \\ 0&0&0 \\ 1&0&0\emat,\quad W_1 =  \bmat 0&0&0 \\ 1&0&0 \\ 0&-1&0 \emat,\quad  W_0 = {2\over 3}\bmat 1&0&0 \\ 0&-2&0 \\ 0&0&1\emat~, \cr & &\cr
W_{-1} &= &\bmat 0&-2&0 \\ 0&0&2 \\ 0&0&0\emat,\quad W_{-2} = 2\bmat 0&0&4 \\ 0&0&0 \\ 0&0&0\emat~. \label{sl3gens}
\eea
The quadratic traces are 
\bea
{\rm tr}_f ( L_0 L_0) &=& 2~,\quad   {\rm tr}_f ( L_1 L_{-1} ) = -4 ~,\cr   {\rm tr}_f ( W_0 W_0 )  &= &{8 \over 3}~,\quad {\rm tr}_f ( W_1 W_{-1} )  = -4 ~,\quad {\rm tr}_f ( W_2 W_{-2} )  = 16~,
\eea
and the non-vanishing symmetric cubic combinations are
\bea
{\rm tr}_f ( L_{(1} L_1 W_{-2)})&=& 8~,\quad {\rm tr}_f ( L_{(1} L_{0} W_{-1)})= -2~,\quad {\rm tr}_f ( L_{(1} L_{-1} W_{0)})= {4\over3}~,\cr
{\rm tr}_f ( L_{(0} L_0 W_{0)})&=& {4\over3 }~,\quad {\rm tr}_f ( L_{(0} L_{-1} W_{1)})= -2~,\quad {\rm tr}_f ( L_{(-1} L_{-1} W_{2)})= 8~,\cr
{\rm tr}_f ( W_{(1} W_1 W_{-2)})&=& -8~,\quad {\rm tr}_f ( W_{(-1} W_{-1} W_{2)})= -8~,\quad {\rm tr}_f ( W_{(1} W_{0} W_{-1)})= {4\over 3}~,\cr
{\rm tr}_f ( W_{(-2} W_0 W_{2)})&=& {32\over 3}~,\quad {\rm tr}_f ( W_{(0} W_{0} W_{0)})= -{16\over 9}~.
\eea
 
As a consequence of \eqref{eq:x33}, in the fundamental representations one can show that 
 \bea
 h_{abc}T^aP^bP^c=P^2-{c_2\over 3} {\bf 1}_{3\times3}~,
 \eea
where $P=P^aT_a$ and $c_2=P^aP^b\delta_{ab}$. 

Using \eqref{charpol3}, the eigenvalues of any element of the algebra $X=X_aT^a$ are
\be
\lambda_X=\sqrt{2 x_2\over 3}\mbox{ diag}\le(\cos\le({x\over 3}\ri),-\cos\le({x\over 3}+{\pi\over 3}\ri) ,-\cos\le({x\over 3}-{\pi\over 3}\ri) \ri)~,
\ee
where 
\be
\cos x \equiv  \sqrt{6} {x_3\over x_2^{3/2}}~, \quad x_2=X^aX^b\delta_{ab}~, \quad x_3= X^aX^bX^ch_{abc}~.
\ee
\section{Alternative derivation of thermal entropy} \label{app:entropy}

In this appendix we give an alternative derivation to the thermal entropy in section \ref{sec:thermhs}. The final result will completely agree with \eqref{entexp}; however we will see that this approach makes it rather clear how our analysis relates to the results reported in \cite{Perez:2012cf,Campoleoni:2012hp,Perez:2013xi}.

 The equations of motion for the massive probe are given by \eqref{first3}. We can recast these equations as
\bea
\bar A_s-A^u_s = 2 \lambda_2 P + 3\lambda_3 h_{abc}T^aP^bP^c ~, \label{eq:eom11}\\
 \frac{dP}{ds} + [\bA_s, P]=0~, \label{eq:eom12}
\eea
where $A^{u}_s = U^{-1} \le(A_s + \frac{d}{ds}\ri)U$. 

For $c_3=0$, we know that the on-shell action  \eqref{eq:3act} depends solely on the combination $\lam_2(s) c_2$. This can be solved in complete generality as follows. Define
\be
A_2\equiv {\rm Tr}(\bar A_s-A^u_s)^2~, \quad A_3\equiv {\rm Tr}(\bar A_s-A^u_s)^3~, 
\ee
with `$\Tr$' given by \eqref{eq:deftr}.  It is very tempting to identify $A_2$ and $A_3$ with the metric-like fields $g_{\mu\nu}$ and $\phi_{\mu\nu\rho}$ in \eqref{eq:metricfield}; this would be the case if $U(s)=1$, however this choice is not generically compatible with the equations of motion for the reasons discussed in section \ref{sec:lack}. Then \eqref{eq:eom11} implies
\bea\label{eq:A2A3}
A_2 &=& 4\lambda_2^2c_2  +{3\over 2}\lambda_3^2 (c_2)^2~,\cr
A_3 &= &6\lambda_2^2\lambda_3(c_2)^2-{3\over 4}\lambda_3^3 (c_2)^3~.
\eea
By combining both equations, we can eliminate $\lambda_3$ and obtain the following equation for $\lambda_2$
\be\label{eq:L3}
\left(16\lambda_2^2c_2-A_2 \right)^2\left(A_2 - 4\lambda_2^2c_2\right)=9 A_3~.
\ee
This equation has 6 roots which we denote $\pm\lambda_{2,i}$. 
The solutions to \eqref{eq:L3} are given by
\bea\label{eq:6roots}
16\lambda_{2,1}^2 c_2 &=& 2A_2 - (A_2)^2 H^{-1/3} -H^{1/3} ~,\cr
16\lambda_{2,2}^2 c_2 &=& 2A_2 +{1+i\sqrt{3}\over 2}(A_2)^2 H^{-1/3}+{1-i\sqrt{3}\over 2}(A_2)^2 H^{1/3}~,\cr
16\lambda_{2,3}^2 c_2 &=& 2A_2 +{1-i\sqrt{3}\over 2}(A_2)^2 H^{-1/3}+{1+i\sqrt{3}\over 2}(A_2)^2 H^{1/3}~,
\eea
with
\be
H\equiv \left(3A_3 +\sqrt{9(A_3)^2 -(A_2)^3}\right)^2~.
\ee
Depending on $A_2$ and $A_3$ not all roots are real. We can write this is in a more compact way. Define
\be\label{eq:angle}
\cos\psi(s) \equiv  {3A_3\over A_2^{3/2}} ~,
\ee
which gives $H=(A_2)^3 e^{2i\psi}$; note that the ``angle'' $\psi(s)$ is not necessarily real. In terms of $\psi$, \eqref{eq:6roots} reduces to
\bea\label{eq:6rootsx}
\lambda_{2,1}^2 = {A_2\over 4c_2}\sin^2\le({\psi\over 3}\ri)~,\quad  \lambda_{2,2}^2 = {A_2\over 4c_2}\cos^2\le({\psi\over 3}-{\pi\over 6}\ri)~,\quad  \lambda_{2,3}^2 = {A_2\over 4c_2}\cos^2\le({\psi\over 3}+{\pi\over 6}\ri)~.
\eea
At this stage it useful to compare with the analysis in the main text. In the notation used in section \ref{sec:thermhs} each of the roots \eqref{eq:6rootsx} represents different parings of the eigenvalues when solving \eqref{eq:matcheigen}. Also, in close analogy to the results in section \ref{sec:entDE}, each root will map to the possible charges the probe can carry given the constraint $c_3=0$. In the notation used here it is rather cumbersome  to compute the charges carried by $P$; the techniques used in the main sections allow for a much more clear analysis of the probe.  

The result \eqref{eq:6roots} is not the end of the story; we still need to impose \eqref{eq:eom12} and the boundary conditions on the path and probe. In this appendix we will consider only the case of closed loops, in which case the probe has to satisfy
\be
U(s_f)=U(0)~,\quad P(s_f)=P(0)~.
\ee
A solution that satisfies the periodicity condition is to take $P=P_0$ (a constant element), and due to \eqref{eq:eom12} we have $[P,\bar A_s]=0$.  If $P$ commutes with $\bar A_s$, then \eqref{eq:eom11} will imply  as well 
\be
[A^u_s,\bar A_s]=0~.
\ee
This imposes a constraint on $U$ very similar to the condition we found for $u_0$ which gives \eqref{eq:matcheigen}. And it should be clear that $U=1$ is not adequate, since in general $A$ and $\bar A$ do not commute. 

Given these additional constraints, we still have to determine $\lambda_2$ as a function of $s$. This information is contained in \eqref{eq:eom11}, but from the way we set the problem in this appendix it requires to solve as well for the path $x^\mu(s)$ that minimizes the action.   This approach  definitely obscures the topological nature of the probe.  Nevertheless,  it is rather interesting that our results here have a close relationship to the expressions obtained in \cite{Perez:2012cf,Campoleoni:2012hp,Perez:2013xi}.  The appeal of expressions like \eqref{eq:angle} is that it gives a hint towards which combinations of metric-like fields are the appropriate invariants to cast the theory in terms of generalized diffeomorphism as defined in \cite{Campoleoni:2012hp}.\footnote{We thank A. Campoleoni for emphasizing this aspect of the computation to us.} 

\section{Computations for principal embedding black hole} \label{app:princ}

Here we present some details of the computations for the principal embedding black hole; essentially we need to compute traces of the matrix $M$, defined by \eqref{Mdef}:
\be
M = \exp(-2\rho_0 L_0)\exp\le(\Delta\phi a_{\phi}\ri)\exp(2\rho_0 L_0) \exp(-\Delta\phi \ba_{\phi}), \label{Mapp}
\ee
where the connections $a, \ba$ are
\begin{align}
a & = \le(L_1 - \frac{2\pi}{k} \sL L_{-1} - \frac{\pi}{2k} \sW  W_{-2}\ri) dx^+ + \mu\le(W_2 - \frac{4\pi \sL}{k} W_0 + \frac{4\pi^2 \sL^2}{k^2}W_{-2} + \frac{4\pi \sW}{k}L_{-1}\ri) dx^- \\
\ba & = -\le(L_{-1} - \frac{2\pi}{k} \sL L_{1} + \frac{\pi}{2k} \sW W_{2}\ri) dx^-  + \mu\le(W_{-2} - \frac{4\pi \sL}{k} W_0 + \frac{4\pi^2 \sL^2}{k^2}W_{2} - \frac{4\pi \sW}{k} L_{1}\ri) dx^+ \label{princBH2},
\end{align}
and where $\sW$ and $\mu$ can be related to a dimensionless parameter $C$ as
\be
\sW = \frac{4(C-1)}{C^{3/2}} \sL \sqrt{\frac{2\pi \sL}{k}}~, \quad \mu = \frac{3\sqrt{C}}{4(2C - 3)}\sqrt{\frac{k}{2\pi \sL}}~,\quad {\mu \over \beta}={3\over 4\pi}{(C-3)\sqrt{4C-3}\over (3-2C)^2} \ .
\ee
To evaluate the matrix exponentials in \eqref{Mapp} it is most convenient to diagonalize $a_{\phi}$ and $\ba_{\phi}$: 
\be
a_{\phi} = V_a \lam_{\phi} V_a^{-1} \qquad \ba_{\phi} = V_{\ba} \bar{\lam}_{\phi} V_{\ba}^{-1}
\ee
where the eigenvalues and eigenvectors may be explicitly computed:
\begin{align}\label{eq:appL}
\lam_{\phi} & = 2\sqrt{\frac{2 \pi \sL}{k}} \mbox{diag}\le(\frac{3 + C\le(\sqrt{4 C - 3} - 2\ri)}{\sqrt{C}(2 C - 3)},\frac{2}{\sqrt{C}}, \frac{3 - C\le(2 + \sqrt{4C - 3}\ri)}{\sqrt{C}(2 C - 3)} \ri) \\
V_a & = \left(
\begin{array}{ccc}
 \frac{2}{C} \left(C-1-\sqrt{4 C-3}\right) & \frac{2 }{C} (2-C) & \frac{2}{C} \left(C-1+\sqrt{4 C-3}\right) \\
 \left(\sqrt{4 C-3}-1\right) \sqrt{\frac{k}{2 \pi {\cal L} C}} &2  \sqrt{\frac{k}{2 \pi {\cal L} C}}  & - \left(\sqrt{4 C-3}+1\right) \sqrt{\frac{k}{2 \pi {\cal L} C}} \\
 \frac{k}{2  \pi {\cal L} } & \frac{k}{2  \pi {\cal L} } & \frac{k}{2  \pi {\cal L} } \\
\end{array}
\right)
\end{align}
and the corresponding objects for the barred connection may easily be constructed:
\be
\bar{\lam}_{\phi} = -\lam_{\phi} \qquad V_{\ba} = g V_{a}
\ee
where the matrix $g$ is
\be
g = \le(\begin{array}{ccc}
0 & 0 & -2 \\
0 & -1 & 0 \\
-\ha & 0 & 0
\end{array}\ri)
\ee
Recall now that we would like to compute $m_1$ and $m_2$, which are defined to be \eqref{m1m2def}
\be
{\rm tr}_f(M) = \frac{m_1}{\ep^{4}} + \sO(\ep^{-2})\qquad {\rm tr}_f(M)^2 - {\rm tr}_f(M^2) = \frac{2 m_2}{\ep^{4}} + \sO(\ep^{-2}) 
\ee
Working out the matrix exponentials as $e^{\Delta\phi a_{\phi}} = V_a e^{\Delta\phi \lam_{\phi}} V_a^{-1}$ and similarly for the barred connection we can compute $m_1$ and $m_2$.  We find

\begin{align}\label{eq:m1}
m_1 &= \le(\frac{C}{4 x (-3+C) } \ri)^2\left[-2xe^{\lam_2\Delta\phi}+e^{\lam_3\Delta\phi}(-3+x)+e^{\lam_1\Delta\phi}(3+x)\right]^2 \le({k\over 2\pi \sL}\ri)^2~, \cr
m_2
&= {1\over 2}\le(\frac{Ce^{\lam_2\Delta\phi}}{2 x (-3+C) } \ri)^2 \left[-2xe^{-2\lam_2\Delta\phi}+e^{\lam_1\Delta\phi}(-3+x)+e^{\lam_3\Delta\phi}(3+x)\right]^2\le({k\over 2\pi \sL}\ri)^2~,
\end{align}
where we defined the diagonal entries in \eqref{eq:appL} as
\be
\lam_{\phi} \equiv \mbox{diag}\le(\lam_1,\lam_2,\lam_3\ri)~,
\ee 
and we also defined $x\equiv \sqrt{-3+4 C}$. 

Besides the limits implemented in section \ref{sec:openPEBH}, we can also consider the two following limits: $\mu$ fixed while $\beta\to 0$, and  $\beta$ fixed while $\mu\to 0$. Using \eqref{eq:m1} and \eqref{eem1m2} we find
\be
S_{\rm EE}(\beta\to 0, \mu\, {\rm fixed})= {c_P\over 3} \log\left(  {\Delta\phi \over \epsilon} \left|1-{16\mu^2 \over \Delta\phi^2}\right|^{1/4}\right)+\ldots
\ee
and
\bea
&&S_{\rm EE}(\mu\to 0, \beta\, {\rm fixed})= {c_P\over 3}\log\left({\beta\over\pi \epsilon} \sinh\left({\pi \Delta \phi\over \beta}\right)\right) \cr &&+ {c_P\over 18}\frac{\pi^2 \mu^2}{\beta ^2}\sinh^{-4}\left(\frac{\pi  \phi }{\beta }\right) \Bigg[8 \left(1-3 \left({2\pi \Delta \phi\over \beta}\right)^{2}\right) \cosh\left(\frac{2 \pi  \Delta\phi }{\beta }\right)\cr &&-3  -5   \cosh\left(\frac{4 \pi  \Delta\phi }{\beta }\right)+{16\pi \Delta \phi\over \beta} \left(\sinh\left(\frac{2 \pi \Delta \phi }{\beta }\right)+\sinh\left(\frac{4 \pi \Delta \phi }{\beta }\right)\right)\Bigg] +\ldots \cr &&
= {c_P\over 3}\log\left({\beta\over\pi \epsilon} \sinh\left({\pi \Delta \phi\over \beta}\right)\right) \cr &&+ {c_P\over 18}\frac{4\pi^2 \mu^2}{\beta ^2}\sinh^{-4}\left(\frac{\pi  \phi }{\beta }\right) \Bigg[-\sinh^2\left(\frac{\pi \Delta \phi }{\beta }\right)\left(1+5\cosh\left(\frac{2 \pi \Delta \phi }{\beta }\right)\right)  \cr &&+{4\pi \Delta \phi\over \beta} \sinh\left(\frac{2 \pi \Delta \phi }{\beta }\right) \left(1 +2\cosh\left(\frac{2 \pi \Delta \phi }{\beta }\right)\right)-6 \left({2\pi \Delta \phi\over \beta}\right)^{2} \cosh\left(\frac{2 \pi  \Delta\phi }{\beta }\right)\Bigg] +\ldots
\eea

\section{Coordinate representation of probe}\label{app:carlip}
In this appendix we review the construction of  $W_{\cal R}(C)$ for infinite dimensional representations originally performed in \cite{Witten:1988hf,Carlip:1989nz} for three dimensional gravity. This approach makes  evident rather quickly a more geometrical interpretation of  $W_{\cal R}(C)$, but is somewhat difficult to generalize to higher spin theories. 
 
Consider first the quantum system with action
\be
S(p,q)_{C,{\rm free}} = \int_C ds\le( p_{\mu}\frac{dq^{\mu} }{ds}+ \lam(s)\le(p_{\mu}p_{\nu}g^{\mu\nu}(q) - m^2\ri)\ri)~. \label{globalac}
\ee
Here $q^{\mu}$ is a set of coordinates on AdS$_3$ and $p_{\mu}$ are their conjugate momenta. $g_{\mu\nu}(q)$ is the pullback metric of AdS$_3$ expressed as a function of the $q^{\mu}$ -- we will see why this is necessary shortly. This AdS$_3$ is the same as the group manifold of $\slt$ discussed in the main text. 
$\lam(s)$ is a Lagrange multiplier that enforces a mass-shell condition. We have as well the Poisson bracket structure
\be
\{p_{\al},q^{\beta}\} = -\delta^{\beta}_{\al}~.
\ee

Note as an aside that if we integrate out $p_{\mu}$ we find
\be
S_{C,{\rm free}}= \int_C ds\le(-\frac{1}{4\lam(s)}\dot{q}^{\mu}\dot{q}^{\nu}g_{\mu\nu}(q) - \lam(s)m^2\ri)
\ee
Further integrating out $\lam(s)$ we find
\be
S_{C,{\rm free}} = -m \int_C ds \sqrt{\dot{q}^{\mu}\dot{q}^{\nu}g_{\mu\nu}(q)} \label{oldac}
\ee
which is indeed the classical worldline length in AdS$_3$.\footnote{At this stage there is actually a simple way to cast the discussion without introducing an auxiliary manifold $g_{\mu\nu}(q)$. To bring this case to the more general setting discussed in section \ref{sec:ctp}, recall that the metric of a group manifold can be written as 
$$
g_{\mu\nu}(q)= \Tr \left(U^{-1}\partial_\mu U U^{-1}\partial_\nu U \right)~,\quad \partial_\mu \equiv  {\partial\over \partial q^\mu}~.
$$
 It is straightforward from here to reproduce \eqref{ac1}.} 

We will first understand the global $SO(2,2)$ symmetry structure for the probe.   Consider the set of AdS$_3$ isometries $(\xi^{\mu}_a, \bar{\xi}^{\mu}_a)$, where the $\xi$ generate $\slt_L$ and the $\bar{\xi}$ generate $\slt_R$. Viewed as vector fields they are functions of the $q^{\mu}$, and  their Lie brackets satisfy the $\slt$ algebra
\be
\sL_{\xi_a}\xi_b^{\mu} \equiv \le(\frac{\p\xi_b^{\mu}}{\p q^{\sig}}\xi^{\sig}_a - \frac{\p{\xi_a^{\mu}}}{\p q^{\sig}}\xi^{\sig}_b\ri) = \epsilon_{abc}\xi^{\mu}_c~. \label{lie}
\ee
A similar relation holds for $\bar \xi$;  further we have $\sL_{\xi_a}\bar \xi_b^{\mu}=0$.  The symmetries of the worldline action \eqref{globalac} are generated by
\be
\sJ^+_{a} \equiv \xi^{\mu}_a(q)p_{\mu}~,\quad \sJ^-_{a} \equiv \bar\xi^{\mu}_a(q)p_{\mu}~. \label{eq:Jp}
\ee
Computing the Poisson bracket of two $J's$ we have
\be
\{\sJ^+_{a}, \sJ^+_{b}\}  = p_{\mu}\le(\frac{\p\xi_a^{\mu}}{\p q^{\sig}}\xi^{\sig}_b - \frac{\p{\xi_b^{\mu}}}{\p q^{\sig}}\xi^{\sig}_a\ri) = -\epsilon_{bac}p_{\mu}\xi^{\mu}_c = \epsilon_{abc}\sJ^+_c~,
\ee
and thus they realize the $\slt$ algebra on the $p$'s and $q$'s. We can now compute the action of these on the canonical variables via Poisson brackets:
\be
\delta_a q^{\mu} \equiv \{q^{\mu},\sJ_a^+\} = \xi^{\mu}_a~, \quad \delta_a p_{\mu} \equiv \{p^{\mu}, \sJ_a^+\} = -\frac{\p \xi^{\nu}_a}{\p q^{\mu}}p_{\nu}~, \quad \delta_a \xi^{\mu}_b \equiv \{\xi_b^{\mu},\sJ_a^+\} = \frac{\p\xi^{\mu}_b}{\p q^{\al}}\xi^{\al}_a~. \label{global}
\ee
Finally, the action \eqref{globalac} is invariant under this global symmetry provided that
\be
\delta_a \le(p_{\mu}p_{\nu}g^{\mu\nu}(q)\ri) = p_{\mu}p_{\nu}\le(-\frac{\p \xi_a^{\mu}}{\p q^{\al}}g^{\al\nu} - \frac{\p \xi^{\nu}_a}{\p q^{\al}}g^{\mu\al} + \frac{\p g^{\mu\nu}}{\p q^{\sig}} \xi^{\sig}_a\ri) = p_{\mu}p_{\nu} \sL_{\xi_a}(g^{\mu\nu}) = 0~,
\ee
and the analogous condition for $\bar \xi$. Hence the vectors $(\xi,\bar \xi)$ must be the Killing vectors of the induce metric, which implies that the probe is propagating on an AdS$_3$ background.

Now we would like to promote this global symmetry to a gauge symmetry. Recall that this system is embedded in an ambient 3D space that does not have a metric structure and only has CS gauge fields turned on; thus in addition to the coordinate $q^{\mu}(s)$ that is used for representing the $\slt$ algebra, there is an extra 3D space in which the Wilson line lives, parametrized by coordinates $x^i$. Take the curve  $C$ of the Wilson line to be parametrized by $x^i(s)$. We want to allow the gauge transformation to depend on those coordinates, i.e. given a function $\Lambda^a(x)$ on this ambient space we have: 
\be
\delta_a q^{\mu} =  \Lambda^a(x(s))\xi^{\mu}_a~, \quad \delta_a p_{\mu} = -\Lambda^a(x(s))\frac{\p \xi^{\nu}_a}{\p q^{\mu}}p_{\nu} ~.\label{gauge}
\ee
The required generalization is
\bea
S(p,q; A,\bar A)_C 
=\int_C ds\le( p_{\mu}D_s q^{\mu} + \lam(s)\le(p_{\mu}p_{\nu}g^{\mu\nu}(q) - m^2\ri)\ri)~, \label{gaugeac}
\eea
where the gauge-covariant derivative of $q$ is
\be
D_{s}q^{\mu} = \frac{dq^{\mu} }{ds}+ \le(A_{i}^a \xi^\mu_a + \bar{A}_{i}^b\bar{\xi}^{\mu}_b\ri)\frac{dx^i}{ds}~.
\ee
Using \eqref{gauge} and the transformation of the gauge field with gauge parameter $\Lambda=\Lambda^a(x(s))J_a^{\pm}$ we can verify that the action \eqref{gaugeac} is invariant under local gauge transformations. This requires the fact that the Lie derivatives of the vector fields $\xi^{\mu}_a$ generate the correct algebra. Note that this construction actually appears to work for any group provided that we can find some manifold to realize the isometries.  

Writing $S_C=S_{C,\rm free} +S_{C,\rm int}$, the interacting piece is
\bea
S_{C,\rm int}=\int_C ds (A_{i}^a \sJ^+_a + \bar{A}_{i}^b \sJ^-_b)\frac{dx^i}{ds}= \int_C {\cal A}~.
\eea
Hence the path integral of the probe is simply
\bea
 \int [ \sD p \sD q \sD \lam] \exp(S(p,q; A,\bar A)_C) = \langle f |   {\cal P}\exp \int_C {\cal A} |i \rangle 
\eea
where the initial and final states  $|i\rangle$  and $ |f\rangle$ contain the boundary conditions of the probe at the end point of the curve $C$. This wave function may be viewed as matrix elements of the Wilson line operator in an infinite-dimensional representation. The mass, spin and conserved charges of the probe---given by $\sJ^{\pm}_a$ in \eqref{eq:Jp}---label the representation $\cal R$.

Now integrating out $p(s)$ and $\lam(s)$ as before we find
\be
S(q; A,\bar A)_C  = m\int_C ds \sqrt{D_s q^{\mu}(s) D_s q^{\nu}(s) g_{\mu\nu}(q)} 
\ee
This is the analog of \eqref{gaugednewac}. It suggests that in the large-$m$ and fixed $\cal A$ limit we can approximate the answer by minimizing the action of a point particle $q^{\mu}(s)$, which is moving in an AdS$_3$ background. However this AdS$_3$ is not (in an obvious way, at least) the geometry of the bulk: it is instead simply an auxiliary space that is a device to realize the $\slt$ algebra. The state of the bulk (e.g. whether or not there is a BTZ black hole in it) enters through the dependence on the background $\cal A$, and {\it not} in a change of the metric $g_{\mu\nu}(q)$.


\end{appendix}

 \end{document}